\documentclass[compsoc]{IEEEtran}

\usepackage{amssymb}
\usepackage{graphicx}
\usepackage{url}
\usepackage{balance}

\usepackage{amssymb}
\usepackage{amsmath}
\usepackage{amsfonts}
\usepackage{algorithmic}
\usepackage[linesnumbered,ruled,vlined]{algorithm2e}
\usepackage{ctable}
\usepackage{url}
\usepackage{subcaption}
\usepackage{hyperref}

\title{JobPruner: A Machine Learning Assistant for\\ Exploring Parameter Spaces in\\ HPC Applications\thanks{
Author preprint. Official Version at \href{http://www.sciencedirect.com/science/journal/0167739X}
{Future Generation Computer Systems (FGCS)} DOI: \href{https://doi.org/10.1016/j.future.2018.02.002}{10.1016/j.future.2018.02.002} (Elsevier).}}

\author{Bruno Silva\footnote{Corresponding author: sbruno@br.ibm.com}, Marco A. S. Netto, Renato L. F. Cunha\\IBM Research}

\vspace{-15mm}
\begin{document}

\clearpage\maketitle
\thispagestyle{empty}

\begin{abstract}

High Performance Computing (HPC) applications are essential for scientists and engineers to create and understand models and their properties. These professionals depend on the execution of large sets of computational jobs that explore combinations of parameter values. Avoiding the execution of unnecessary jobs brings not only speed to these experiments, but also reductions in infrastructure usage---particularly important due to the shift of these applications to HPC cloud platforms. Our hypothesis is that data generated by these experiments can help users in identifying such jobs. To address this hypothesis we need to understand the similarity levels among multiple experiments necessary for job elimination decisions and the steps required to automate this process. In this paper we present a study and a machine learning-based tool called JobPruner to support parameter exploration in HPC experiments. The tool was evaluated with three real-world use cases from different domains including seismic analysis and agronomy. We observed the tool reduced 93\% of jobs in a single experiment, while improving quality in most scenarios. In addition, reduction in job executions was possible even considering past experiments with low correlations.

\end{abstract}


%
%


\section{Introduction}

Big data and artificial intelligence are becoming crucial for the development of solutions in various fields including agronomy, health, aerospace, circuit design, astronomy, and oil \& gas exploration. High Performance Computing (HPC) simulations used in these fields produce plenty of data that could be used to optimize new experiments. Such data is becoming even more pervasive due to the increasing efforts in improving reproducibility of computational experiments~\cite{stodden2016enhancing,chard2015globus,santana2017reproducibility}. Natural questions are: what is the actual benefit of exploiting data from previously executed experiments and how to do that properly. This paper attempts to shed some light into answering these questions.

Ongoing efforts to reduce execution time of HPC experiments come mainly from (i) optimization of hardware and software layers that host user applications and (ii) use of Design of Experiments (DOE)~\cite{kleijnen2005state} techniques to avoid running unnecessary jobs or prioritize jobs that provide the best ``return of investment''---by doing so, users can run experiments in a more efficient manner with the same, or even a reduced, amount of computing power~\cite{sanchez2009better}. These reduction costs are particularly important due to the increasing efforts in moving HPC applications to cloud platforms~\cite{netto2015deciding,gupta2013who}.

With the advances in the areas of big data and artificial intelligence, along with HPC platform improvements, mainly from the GPU arena, a complementary direction can be pursued to help users run their scientific experiments faster. Our goal here is to use data produced by previous experiments to help in new ones. We focus on the problem of reducing the number of jobs to explore parameters of user applications, which is a common process in scientific experiments when users have to calibrate models (e.g. crops, airplane wings, and circuit boards) or understand various what-if scenarios (e.g. soil properties, atmosphere conditions, and material properties). Users typically have to study a similar object/model but with different conditions/properties and use different strategies to make exploration vs exploitation decisions. One way to extract such a knowledge is via machine learning, where patterns can be found from previous executions that contain similar characteristics.

In this paper, we provide a step towards answering the question of how much knowledge can be leveraged from previous experiments to accelerate new ones. We analyze previous executions of real-world applications with similar search spaces to assess whether we can generate hints to users on which jobs can be prioritized or canceled on ongoing experiments. This analysis makes use of optimization and machine learning techniques. Our main contributions are therefore:

\begin{itemize}
    \item JobPruner, a tool to help users explore parameter spaces in HPC experiments
by building machine learning models from data obtained from previous executions (Section~\ref{sec:solution});
    \item Understanding of how reusing data from past executions affects results by evaluating three real HPC parametric applications from different fields, including seismic analysis and agriculture. The evaluation measures possible reductions in number of jobs when time for experimentation is limited and the impact on results considering different knowledge base sizes (Section~\ref{sec:evaluation}).
\end{itemize}

\section{Background}
\label{sec:background}

Several research groups have conducted studies and developed technologies to assist users in their large scale HPC experiments. Related to our work are efforts from multiple communities, including computational steering, workflow management, design of experiment, optimization, and more recently big data and artificial intelligence. In this section, we describe some of these efforts and a formal description of the problem tackled in this paper.

\subsection{Related Work}
\label{sec:relatedwork}

Computational steering \cite{mulder1999survey,wright2010steering,garcia2015computational,danani2015computational,kail2015novel} aims at providing users with tools for parameter reconfiguration of ongoing experiments. Parker and Johnson \cite{parker1995scirun} introduced a system called SCIRun that uses a dataflow programming model and visual programming to simplify the tasks of creating, debugging, optimizing, and controlling complex scientific simulations. Van Wijk \textit{et al.}~\cite{van1997bringing} highlighted that the implementation of computational steering in practice is hard. To overcome this problem, they implemented an environment in which a data manager facilitates the interaction between user applications and steering components. Chin \textit{et al.}~\cite{chin2003steering} incorporated computational steering in mesoscale lattice Boltzmann simulations and showed the benefits of their work. They discussed that large scale simulations require not only computational resources but tools to manage these simulations and their produced results, what they called \textit{simulation-analysis loop}. Netto \textit{et al.}~\cite{netto2005scheduling} introduced a scheduler system to automatically offer more resources to parametric application jobs based on the quality of their intermediate generated results so as users could get faster to their desired goal. More recently, Mattoso \textit{et al.}~\cite{mattoso2015dynamic} surveyed the use of steering in the context of HPC scientific workflows highlighting a tighter integration between the user and the underlying workflow execution system.

Workflow management systems \cite{ludascher2006scientific,deelman2009workflows,gil2011wings,deelman2017future} are also relevant to help users in complex scientific experiments. Deelman \emph{et al.}~\cite{deelman2009workflows} developed a taxonomy of e-Science systems so scientists can assess the suitability of workflow systems to their experiments. Gil \emph{et al.}~\cite{gil2011wings} introduced Wings, a system that uses workflows to represent computational experiments and reasons about the experiment design space. In doing so, Wings can assist scientists in designing such experiments by tracking constraints and ruling out invalid designs.

Running complex experiments can always benefit from user input, therefore a community of researchers have been working on optimization assisted by humans \cite{meignan2015review,meignan2013interactive}. Meignan \textit{et al.}~\cite{meignan2015review} provided a detailed survey and taxonomy of efforts in interactive optimization applied to operations research. They explored the different roles a user can have in an optimization process, such as adjusting or adding new constraints and objective, helping on the optimization process itself, and guiding the optimization process by providing information related to decision variables. Meignan and Knust~\cite{meignan2013interactive} proposed a system that employs user feedback on the optimization to use as long-term preferences for future runs. Nascimento and Eades~\cite{nascimento2005user} proposed a framework for humans to assist the optimization process via inserting domain knowledge, escaping from local minima, reducing the search space to be explored, and avoiding ambiguity for optimal multi-solutions. Abramson \textit{et al.}~\cite{abramson2001automatic} developed a system for users to parameterize arbitrary problems with the goal of optimizing a given objective function. Silva \textit{et al.}~\cite{silva2016sla} introduced a tool that incorporates user hints based on intermediate results to advise them about their strategies when running parametric applications tied to SLA constraints. Luszczek \emph{et al.}~\cite{luszczek2016search} described a system to generate and prune search spaces for auto-tuning matrix multiplication kernels. Their solution is based on a declarative language to describe the search space alongside pruning constraints. Researchers have also explored visualization techniques to add humans to the optimization process \cite{klau2010human,colgan1995cockpit,endert2014human,tweedie1994attribute}. WorkWays \cite{nguyen2015workways,nguyen2012workways} is a science gateway with human-in-the-loop support for running and managing scientific workflows.

The area of Design of Experiments also has contributions to accelerate the evaluation of search spaces. Kleijnen \textit{et al.}~\cite{kleijnen2005state} developed a survey and a user guide on advances in this area until 2005, including benefits and drawbacks of various designs. Sanchez and Wan~\cite{sanchez2009better} also discussed the benefits of design of experiments to avoid brute-force approaches. They also presented a tutorial on how to apply experimental design to simulation experiments. Using the fractional factorial design technique, Abramson \textit{et al.}~\cite{abramson2011parameter} developed a system to facilitate parameter exploration and abstract the underlying computing platform.

The execution of large number of independent jobs may be complex to users, mainly when distributed resources are used to run them. Resources may have different computational power and availability and may have different mechanisms to be accessed. Therefore, over the years, several tools have been created such as Condor \cite{litzkow1988condor}, XtremWeb \cite{fedak2001xtremweb}, BOINC \cite{anderson2004boinc}, Nimrod \cite{abramson2000nimrod}, and OurGrid \cite{andrade2003ourgrid} to facilitate user access to these computational platforms. These tools mainly focus on managing users jobs looking into the computational infrastructure. Our tool aims at looking into the user workload and finding patterns from past experiments, which allow users to drastically reduce their search spaces when performing experiments of similar nature.

With advances in artificial intelligence~\cite{segev2016learn,duan2012domain} and big data~\cite{gao2008knowledge}, researchers are studying how to apply technologies from these areas to optimize the execution of experiments. For instance, in the data mining domain, Padillo \emph{et al.}~\cite{padillo2016subgroup} proposed exhaustive search algorithms for subgroup discovery: identifying relations between a target variable and independent variables. Their algorithms can prune search spaces and find relationships in massive datasets. Weiss \emph{et al.}~\cite{weiss2016survey} built upon the survey conducted by Pan and Yang~\cite{pan2010survey} and investigated papers and software motived by the improvement of target predictive functions from one domain by using predictive functions learned for other domains (i.e. \emph{transfer learning}).

Our research is based on existing efforts in optimization, design of experiments, and machine learning in order to build a tool for reducing the execution time of HPC experiments. Our tool is based on the hypothesis that it is possible to reuse knowledge from previously executed experiments to run new ones. To the best of our knowledge, there is no previous work that leverages previous HPC application experiments to prune search space of new ones.

\subsection{Problem Description}
\label{sec:problem}

Problems from several industries are tackled by users running complex computer simulations and optimizers that constitute a \textit{software} with a set of \textit{parameters}, where each parameter can receive different \textit{values}. The optimization process for a parametric application is named \textit{experiment}, and each individual software execution is a \textit{job}. These jobs can be computational intensive, which may require High Performance Computing (HPC) platforms not only due to processing needs but also due to memory constraints. Although jobs run independently, which is a characteristic known in High Throughput Computing (HTC) applications, each job can run on a multi-core machine or even be distributed in a computing cluster as long as the instruction to do so is available to our tool.

We consider applications with $n$ parameters and each parameter has a finite discrete domain ${\cal D}_i$ where $i \in \{1, \cdots, n\}$. A function $f: {\cal D}^n \rightarrow \mathbb{R}$ is adopted to evaluate the quality of parameter values and is specific for each parametric application. The function may also be subjected to constraints ($t_j \forall j \in \{1, \ldots, m\}$, $u_k \forall k \in \{1, \ldots, p\}$) depending on the problem characteristics. The considered optimization problem $P$ is described as follows:
\begin{align}
      &\underset{\vec{x} \in {\cal D}^n}{\operatorname{maximize}}& & f(\vec{x}) \\
  &\operatorname{subject\;to}
      & &t_j(\vec{x}) \leq 0, \quad j = 1,\dots,m \\
      &&&u_k(\vec{x}) = 0, \quad k = 1, \dots,p
  \end{align}


 We assume the user is able to generate $f(\vec{x})$ from the job output. For instance, if the output $g(\vec{x})$ of a simulation process is a real value that should be minimized, then $f(\vec{x})$ can be assigned as $f(\vec{x}) = -g(\vec{x})$ and the problem definition remains the same. If the parametric application $a$ generates multiple outputs (e.g., $a: {\cal D}^n \rightarrow {\cal O}^m$), a wrapper function $h: {\cal O}^m \rightarrow \mathbb{R}$ should be employed to represent the behavior of $f$ (i.e., $f(\vec{x}) = h(a(\vec{x}))$).

For instance, one of our case study applications (Section~\ref{sec:evaluation}) predicts the production of multiple crops from a single region. We use a wrapper function that sums the crop yields to generate a single objective function. Multiple objective optimization is out of the scope of this paper.



Running all possible values for each parameter can be unfeasible---even with large HPC machines. Therefore, users need strategies to select a subset of values for each parameter that covers parts of the exploration space and gives them enough information to answer their questions. For instance, the user may want to know  what is the best set of parameter values that provides a close-to-optimal solution. The strategies for parameter-value selections can be defined by Derivative Free Optimization (DFO) methods \cite{conn2009introduction}, which correspond to a subject of mathematical optimization where the derivative information is unavailable or impractical to be obtained.

An effective approach to speed up the solution finding in optimization problems corresponds to the reduction of the search space by pruning the range of parameter values or fixing a parameter with a given value. Often users use their background from previous experiments to determine the parameter ranges and/or prune the search space to get faster results. There has been little progress on employing machine learning techniques for reducing search space in HPC experiments.



\section{System and Algorithms}
\label{sec:solution}

This section describes JobPruner, the proposed solution for eliminating non-promising jobs of HPC experiments. JobPruner is part of a larger system,  Copper~\cite{silva2016sla}, aimed at calibrating models and evaluating what-if scenarios by providing an integrated interface for executing experiments and using machine learning for providing users with insights. JobPruner focuses on improving the quality of experiments with restricted resources by using knowledge from already executed experiments to reduce the solution search space.

\begin{figure}[!t]
        \centering
        \includegraphics[width=\columnwidth]{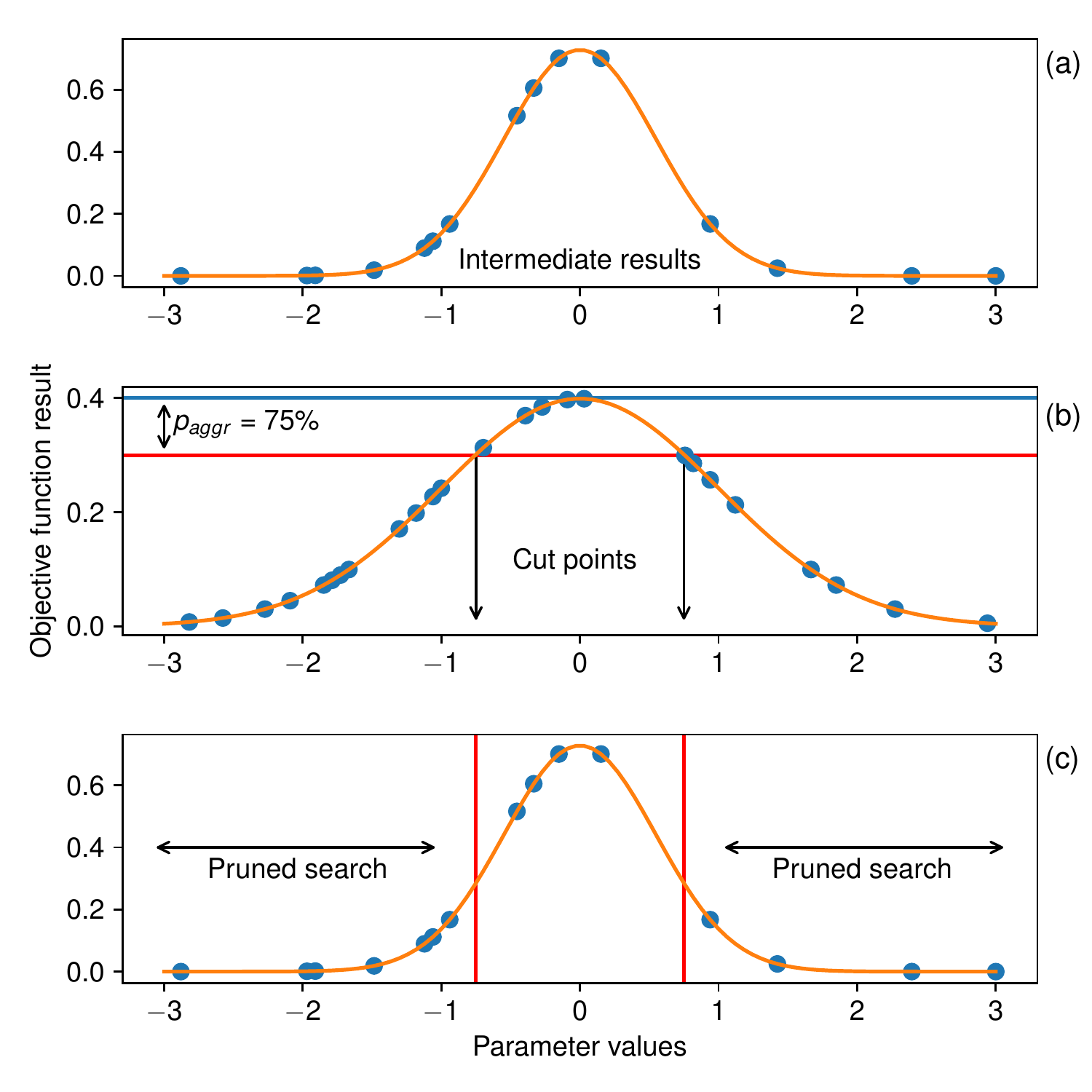}
        \caption{%
            Motivating example of search space reduction using previous
            experiments. In (a) we see a function that will be optimized. In
            (b) we see a function that is used as a template of previous
            evaluations. In (c) we see the function from (a) but with its space
            pruned based on information from (b).
        }\label{fig:motexp}
\end{figure}

\subsection{Motivating Example}

Suppose one wants to maximize a function of a model. Assume the value output by the model depends only on a single independent variable instantiated in a given domain and also that the only way to obtain the output is by executing the model. Also assume this model has been executed for a number of times in different experiments and we are now given the task of evaluating this model in a new setting. For instance, if a crop model for one farm is under evaluation, we use data from previous evaluations on other farms to obtain insights about the current experiment. Figure~\ref{fig:motexp} displays an example in which we want to maximize some function.

The pruning strategy works as follows: initially, the model is evaluated several times and a surrogate function is fit to the obtained results (Figure~\ref{fig:motexp}a). Then, the newly-obtained surrogate is compared to all the previous surrogates found for this model in previous experiments. After the search, we select the previous experiment that has the surrogate closest to the one just obtained. That way, we can exploit the fact of having more evaluations of previous experiments to possibly avoid executing non-promising jobs.

In Figure~\ref{fig:motexp}b, we see such a domain. In our tool, we employ a pruning aggressiveness factor, $p_{aggr}$, that acts as a cutoff parameter: we discard points that generated values smaller than $p_{aggr} \times  \max $, where $\max$ is the maximum value found in the previous experiments. In this example, we use $p_{aggr} = 75\%$ and the maximum value found in that experiment was $0.4$. Therefore, points with value less than $75\% \times 0.4 = 0.3$ are discarded.

\begin{figure}[!t]
        \centering
        \includegraphics[width=\linewidth]{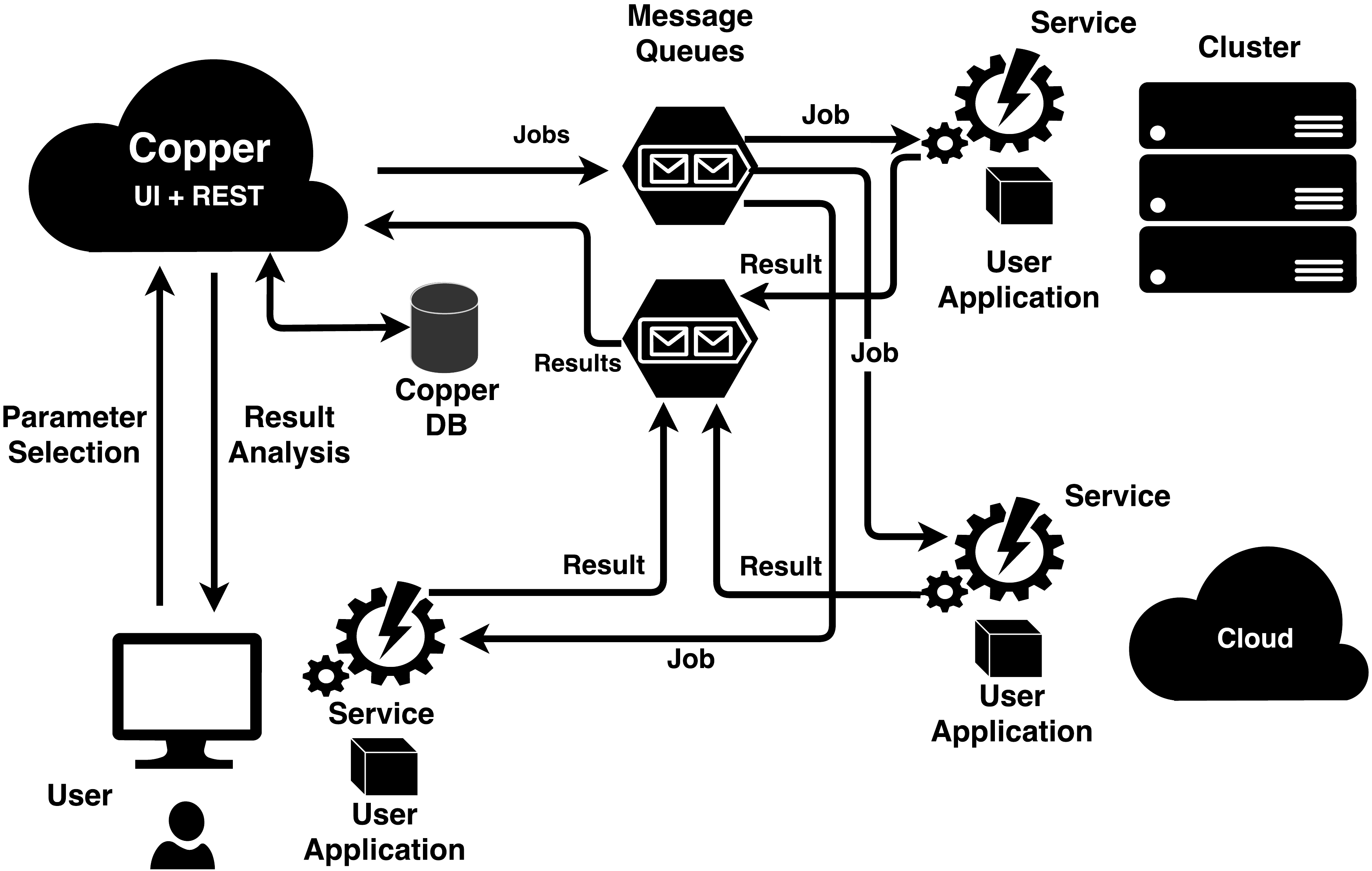}
        \caption{%
            Copper architecture
        }\label{fig:copper_arch}
\end{figure}

The new search space corresponds to parameter values that were not discarded, represented by the points above the red line in Figure~\ref{fig:motexp}b. By using a redefined search space, we may find better results faster for the current experiment, as no computational resources are applied in regions that may lead to non-promising results. Figure~\ref{fig:motexp}c shows the final region to be explored, where only the central region of the search space should be evaluated.

%

\subsection{Copper}

Figure \ref{fig:copper_arch} presents an overview of Copper, which is a cloud service for exploring search spaces in HPC parametric applications. By using Copper, users can prioritize or discard jobs to speedup the evaluation of their applications. The infrastructure is composed of a database (Copper DB) to store jobs results and their respective parameters, a backend (UI + REST), two message queues (Jobs and Results) and a Service for each user application.

The workflow for utilization of copper is the following. First, users select a batch of jobs to be executed, each job contains a single combination of parameters and may return a result (Section~\ref{sec:problem}). Jobs are stored in Copper DB and sent to Jobs message queue (job id and respective parameters). Once user jobs are in the message queue, a service retrieves them for execution. Services can be located in different environments for job execution, including cluster, cloud or even the user personal machine. Once the job finishes, the results are returned to the users via Results message queue. For more details about Copper the reader is referred to \cite{silva2016sla}.

\subsection{JobPruner}
\label{sec:proposed_solution}

Figure~\ref{fig:arch_overview} shows the interaction between Copper and JobPruner and its components. The overall flow between Copper and JobPruner is the following:

\begin{figure*}[!ht]
        \centering
        \includegraphics[width=1.0\linewidth]{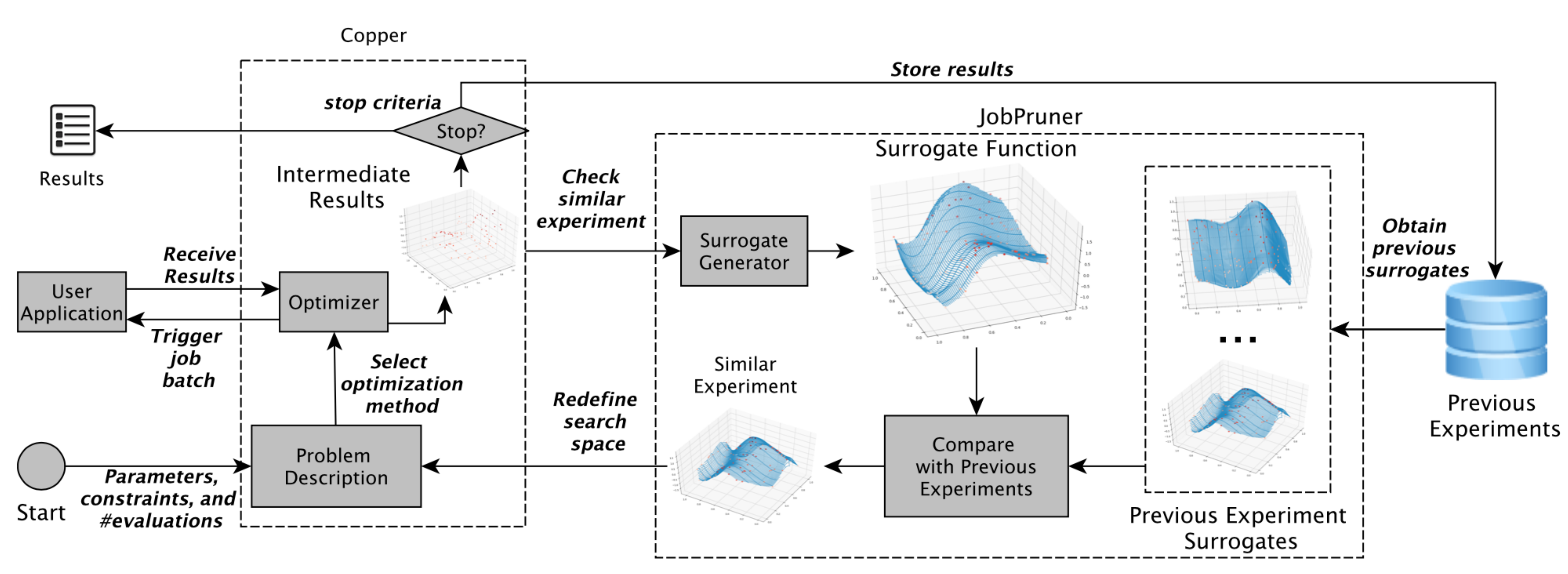}
        \caption{%
            Overview of the proposed pruning approach.
        }\label{fig:arch_overview} 
\end{figure*}

\smallskip

\begin{enumerate}
	\setlength\itemsep{0.5em}
	\item \textbf{Problem description:} having an application registered in Copper, the user specifies a list of parameters to be evaluated, the range of variables each parameter can assume, constraints for time or number of jobs to complete an experiment, an application, and an optimization metric;

	\item \textbf{Optimizer:} With the input provided by the user, the optimizer chooses a set of parameters to be used to evaluate a model input by the user. Copper utilizes DFO strategies such as generalized greedy randomized adaptive search (GRASP), general pattern search (GPS), particle swarm optimization (PSO), and simulated annealing (SA)~\cite{feo1995greedy, conn2009introduction};

	\item \textbf{User Application:} The set of parameters selected by the optimizer is then used for execution of the model. As new results are generated, the Optimizer selects a new batch of parameters to evaluate. Also, as intermediate results become available, the Optimizer communicates with JobPruner to potentially prune the search space;



    \item \textbf{Stop:} The interaction between Optimizer and User Application continues until one stop criterion is met. Examples of criteria are: exceeding the maximum budget of jobs run, exceeding the time allocated for optimization, or successfully optimizing the model.
\end{enumerate}

\medskip
Inside JobPruner, the following steps are executed:
\smallskip

\begin{enumerate}
	\setlength\itemsep{1em}
    \item \textbf{Surrogate Generator:} Copper sends intermediate results to JobPruner to verify if the search space can be reduced. The entry point of JobPruner is the Surrogate Generator, which is responsible for creating a function based on available results of the ongoing experiment. This intermediate surrogate is stored in the Previous Experiments database and is refined each time the surrogate is updated with information from new jobs executed. JobPruner includes the following surrogate generation algorithms: linear, cubic, spline interpolations, and k-nearest neighbors regression. For this work we adopted k-nearest neighbors regression to estimate points that were not evaluated. Other machine learning approaches can be used without loss of generality (e.g., neural networks);

	\item \textbf{Compare with Previous Experiments:} With a surrogate function, it is possible to search the previous experiments database (also known as {\it knowledge base}) for similar surrogates. If a similar surrogate is found, it can be used to reduce the ongoing experiment search space size.

\end{enumerate}

Next we describe how a previous experiment is selected for the pruning process which is the functionally of JobPruner.

\subsection{Previous Experiment Selection}
\label{sec:selection}

JobPruner utilizes the most similar surrogate from previous experiment database to prune the search space. In order to compare two surrogates, we employ normalized cross correlation~\cite{briechle2001template}, a technique from the pattern recognition and machine learning area, which is useful when surrogates with different output ranges are compared. This technique normalizes the outputs by subtracting each output point by the mean and dividing by the standard deviation. As the data is normalized, it is possible to evaluate experiments with different output amplitudes focusing the comparison on data shapes not in specific values. The normalized cross-correlation $(n_{corr})$ between two functions $f(\vec{x})$ and $p(\vec{x})$, with $\vec{x}$ $\in$ ${\cal D}^n$, is described as follows:

\[
    n_{corr} = \frac{1}{s} \sum_{\vec{x}_i}\frac{(f(\vec{x}_i) - \bar{f})(p(\vec{x}_i) - \bar{p})}{\sigma_f\sigma_g},
\]
where $s$ is the number of samples, $\bar{f}$ and $\bar{p}$ are the surrogate mean values of the current experiment and the previous experiment, respectively, $\sigma_f$ and $\sigma_g$ are the surrogate standard deviations for $f$ and $p$. Then, we use the previous experiment with higher $n_{corr}$ to prune the current experiment search space, if $n_{corr}$ is higher than a user-define threshold.

\subsection{Search Space Pruning Algorithm}
\label{sec:pruning_alg}

Algorithm~\ref{alg:prune} is responsible for pruning the search space based on previous experiments. The algorithm receives the previous experiment results set (${\cal P} \subseteq {\cal D}^n \to \mathbb{R}$), the current experiment search space (${\cal C} \subseteq {\cal D}^n$), and the prune aggressiveness ($p_{aggr} \in \mathbb{R}$) as parameters. The model output corresponds to a potentially smaller new search space (${\cal N} \subseteq {\cal C}$).

\begin{algorithm}
\SetAlgoLined%
\SetAlgoNoEnd
\SetKwInOut{Input}{Input}
\SetKwInOut{Output}{Output}
\Input{${\cal P}$, ${\cal C}$, $p_{aggr}$}
\Output{${\cal N}$}
${\cal N}$ $\leftarrow$ ${\cal C}$\;
$cut\_val \leftarrow p_{aggr} \times \max({\cal P})$\;
\ForEach{${\cal D}_i$ of ${\cal N}$}
{%
${\cal D}_t$ $\leftarrow$ ${\cal D}_i$\;
\ForEach{$val$ $\in$ ${\cal D}_i$}
{%
    $\mathit{jobs} \leftarrow \{j \in {\cal P}$ $|$ $\mathrm{pvalue}(j, i) = val\}$\;
    $r \leftarrow \top$\;
    \ForEach{$job \in jobs$}
    {%
        \If {$\mathrm{eval}(job) \ge cut\_val$}{%
            $r \leftarrow \bot$\;
        }
    }
    \If {$r$}{%
        ${\cal D}_t \leftarrow {\cal D}_t - \{val\}$\;
    }
}
${\cal D}_i$ $\leftarrow$ ${\cal D}_t$\;
}
return ${\cal N}$\;
\caption{\emph{prune}, an algorithm for pruning the search space.}
\label{alg:prune}
\end{algorithm}

Initially, ${\cal N}$ receives the search space for the current experiment (Line 1). In Line 2, variable $cut\_val$ is created to establish a threshold for parameters that lead to non-promising results. Lines 3-13 loop over each parameter ${\cal D}_i \in {\cal N}$ to remove parameter values that generate non-promising results. In Line 4, a temporary set ${\cal D}_t$ copies each parameter ${\cal D}_i$. For each parameter value $val$ $\in$ ${\cal D}_i$, we create a variable $jobs$ that corresponds to all jobs from previous experiment ${\cal P}$ that use parameter ${\cal D}_i$ with value $val$ (Line 6). $\mathrm{pvalue}(j \in {\cal P}, i \in \mathbb{N})$ returns the $i$-th parameter value for job $j$. If all values of $job \in jobs$ lead to results below the $cut\_val$ (Lines 7-12), $val$ is removed from ${\cal D}_t$ (Line 12). Otherwise, ${\cal D}_t$ remains with the same values. Parameter ${\cal D}_i$ is updated with the new parameter values set in Line 13. Finally, the new search space is returned in Line 14. JobPruner executes Algorithm~\ref{alg:prune} after each job batch, and updates the current optimization method constraints to generate new jobs only in the newly-obtained search space region.

\subsection{Automatic $p_{aggr}$ Generation}
\label{sec:auto_pagg}

JobPruner can automatically determine prune aggressiveness ($p_{aggr}$) values based on previous experiment data. Higher values of $p_{aggr}$ result in larger cuts in the search space. Therefore, if $p_{aggr}$ gets closer to 1, the probability that an optimal solution will be excluded is higher. Whereas, if JobPruner uses lower $p_{aggr}$ values, it cannot help users speed up their experiments by eliminating non-promising jobs.

Let ${\cal Z}$ be the set of previous experiment outputs (Section \ref{sec:proposed_solution}). We measure the spatial continuity of ${\cal Z}$ using experimental variograms~\cite{puppala2015spatial}. Experimental variograms are defined as one-half of the average squared difference between each pair of points in ${\cal Z}$ with a distance $h$, also known as lag. Formally, we define the experimental variogram $v_r(h)$ as:
\[
    v_r(h) = \frac{1}{2n(h)} \sum_{i=1}^{n(h)} \big[ z(x_i + h) - z(x_i) \big]^2 \text{,}
\]
\noindent where $z(x)$ corresponds to the data output taken at location $x$, $z(x + h)$ is the measurement taken at the point $x + h$, and $n(h)$ the number of data pairs that are $h$ units apart from each other.

An important metric related to experimental variogram analysis corresponds to the variogram \textit{nugget}, which is the value of $v_r(h=0)$. Theoretically, the nugget should be zero, however spatial sources of variation at distances smaller than the sampling interval leads to positive nuggets~\cite{puppala2015spatial}. It can be used as an indicator of how overall data is spatially correlated for small distances \cite{kitanidis1997introduction}. Another metric is the \textit{sill} that defines the variogram limit value which is given by $\lim_{h \to \infty} v_r(h) = variance({\cal Z})$. If we divide the nugget by the sill, we obtain the proportion of data discontinuity for small lags, then the suggested prune aggressiveness $(s_{aggr})$ is given by:

\[
    s_{aggr} = 1 - v_r(0)/variance({\cal Z})
\]

\noindent which indicates how continuous the data is for small lags. In order to use variograms, ${\cal Z}$ should be normally distributed and presents spatially constant mean and variance. JobPruner checks these conditions to make $s_{aggr}$ suggestions, if ${\cal Z}$ does not have these properties then a fixed value (e.g., 0.6) is proposed and the user is notified. We also adopt a user-defined superior limit for $s_{aggr}$ to avoid a too aggressive prune.

\section{Evaluation}
\label{sec:evaluation}

In this section, we evaluate JobPruner to eliminate non-promising jobs in HPC experiments by using three real applications (Section~\ref{sec:solution}). Our hypothesis is that we can use data from previous experiments to help on new ones, and we are interested in understanding the impact of pruning aggressiveness and knowledge base size on the quality of the produced results.

\subsection{Experiment Setup}

We split our evaluation in two sets of experiments using two DFO methods each (PSO and SA).
The first considers the effects of the pruning aggressiveness on the search space reduction and the best solution found (Section \ref{sec:prunaggre}). The second helps us understand the impact of the knowledge base size on the quality of the results (Section \ref{sec:prevdbsize}).

For all experiments, the limit of function evaluations corresponds to 10\% of the search space size. The following metrics are analyzed: (i) difference between the best-found solution and the global optimum, (ii) search space pruning size. We performed all experiments using a 15-node Xeon E5-2680v2 cluster, each node with 20 cores and 128 GB of RAM.

\bigskip

\noindent {\bf Applications.} We selected three applications from different domains and with different search space characteristics. Following is a short description of the applications with an overview of their input parameters and outputs users are interested in.

\smallskip
{\bf 1. Seismic:} A seismic image corresponds to a representation of sound waves through underground rock structures. By using these images, seismologists can estimate the shape and depth of gas and crude oil geological formations. To assist users in this task, we use an in-house application (Seismic) to analyze seismic images using machine learning. The tool has several parameters including a visual descriptor (e.g., local binary patterns) and a classifier (e.g., k-nearest neighbors) to estimate the composition of the different regions of a seismic image. Visual descriptors and classifiers have parameters to be tuned to get proper geological formation estimations. To evaluate the quality of a selection of feature extractor and classifier, the tool employs a database storing reservoir annotated images with a description of each image component (rock layers) for a specific field. Classification performance varies according to the selected feature extractor and classifier and is in the interval [0, 1], where 0 means the classifier gave wrong predictions for all images in validation set and 1 when  all images were classified correctly. The user's goal is to select a visual descriptor  from a set of three, a classifier also from a set of three, and their parameters in order to maximize the classification accuracy of a field seismic image. For each field evaluation (experiment), there are 5502 combinations of classifier and extractor configurations.

\smallskip
{\bf 2. AgroAnalytics:} AgroAnalytics is a tool to help farmers know how they should fertilize their crops. The tool is implemented on top of PCSE (Python Crop Simulation Environment)\footnote{PCSE\@: \texttt{\url{http://pcse.readthedocs.io/en/stable/}}} which corresponds to a python implementation of WOrld FOod STudies (WOFOST) model~\cite{boogaard2014wofost}. As performance metric, we used the yield of winter wheat crops from farms located in the Anhui Province in China. The crops are divided in 50km $\times$ 50km grids and they have their own soil and weather characteristics. The objective of this case study is to maximize the production of the land by choosing how to fertilize the soil. Here the user can select one of four fertilizer configurations in five crop stages. Hence, we may have $4^5=1024$ fertilizing configurations for each crop.

\smallskip
{\bf 3. SchedSim:} Scheduler Simulator (SchedSim) is a tool to assist system administrators to create management policies for HPC clusters. It accepts a variety of parameters including number of processors in the cluster, partitions, and scheduling algorithms. Tuning the scheduler and cluster properties to meet client business goals is not a trivial task and several scenarios must be executed to achieve that. For the evaluation, we had the following input: (i) a workload containing historical data of jobs submitted to a cluster and (ii) five queues, each with a configuration of user requested allocation time. For instance, a queue can only accept jobs with 15 minutes of requested time, whereas another can accept only jobs that take between 2 to 3 hours. We created an array of 13 possibilities of time configurations, thus generating 495 jobs for each experiment (that is, a workload analysis). Having five queues, we used four variables to determine the index in this array of possibilities to generate the ranges of requested times for the five queues. We used a fair share algorithm, given equal share to each queue to focus our analysis only on the parameters of the requested time per queue. The workloads came from a mix of 6-month portions of five clusters from the Parallel Workloads Archive~\cite{feitelson2014experience} (HPC2N, SDSC-SP2, KTH-SP2, SDSC-BLUE, and SDSC-DS)\footnote{Parallel Workloads Archive: \texttt{\url{http://www.cs.huji.ac.il/labs/parallel/workload}}}. The goal in this use case is to know what configuration each queue should have in order to minimize the average slowdown of all jobs in a given workload.

\begin{figure}[!t]
\begin{subfigure}{0.99\linewidth}
        \centering
        \includegraphics[width=0.83\linewidth]{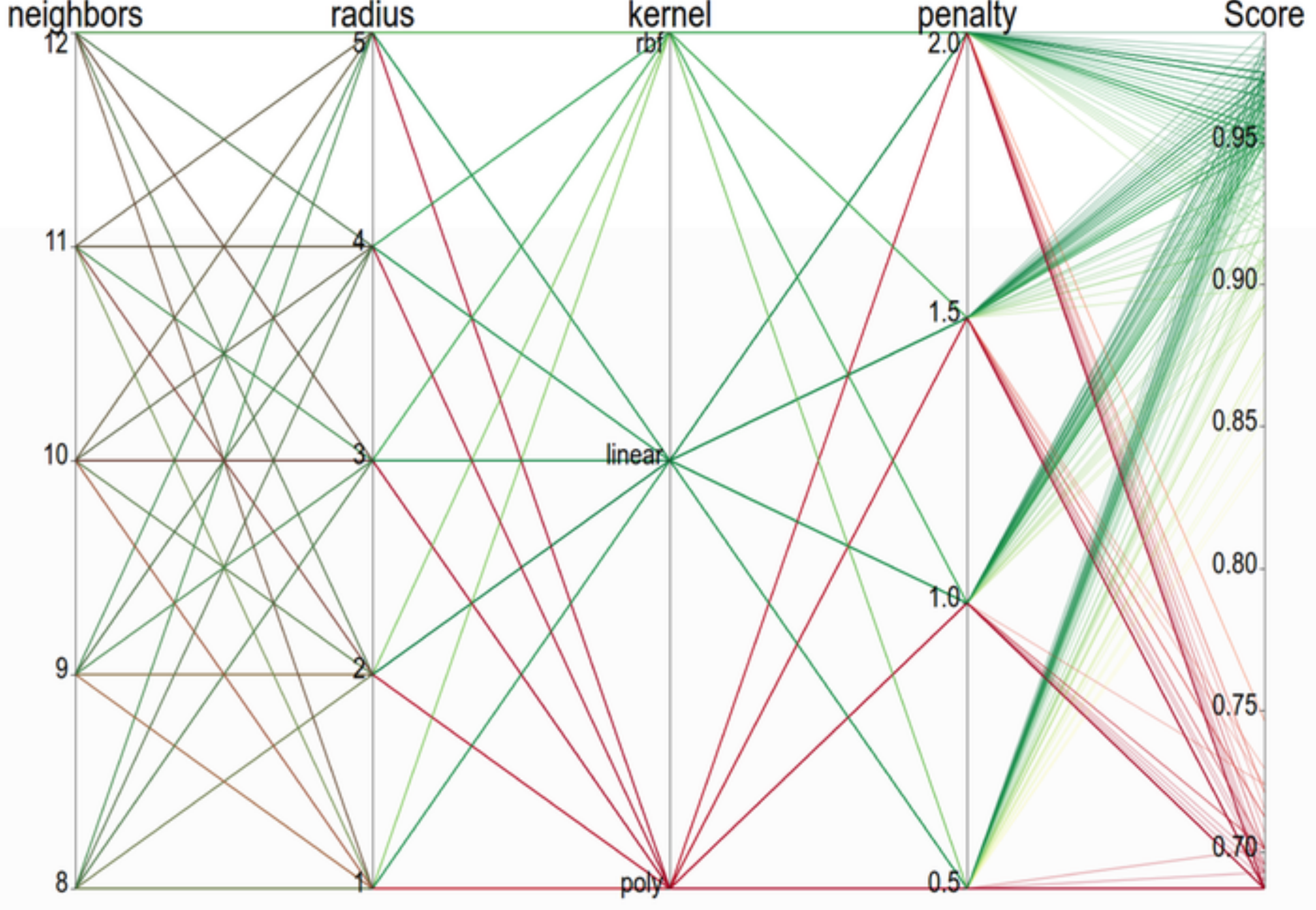}
        \caption{
            Seismic.\vspace{5mm}
        }\label{fig:seismic_ss}
\end{subfigure}
\vspace{3mm}
\begin{subfigure}{0.99\linewidth}
        \centering
        \includegraphics[width=0.83\linewidth]{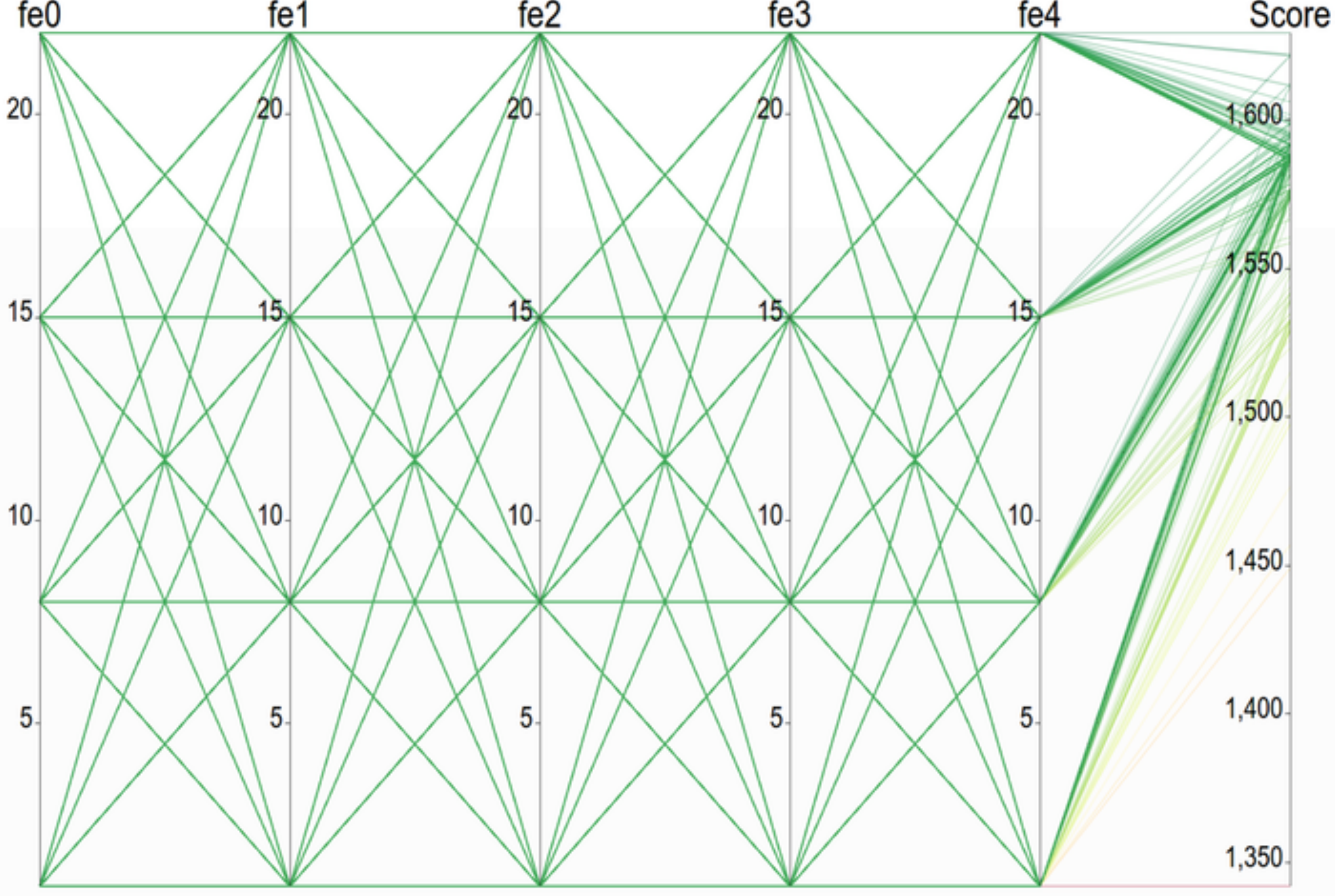}
        \caption{
            AgroAnalytics.\vspace{5mm}
        }\label{fig:agro_ss}
\end{subfigure}
\vspace{3mm}
\begin{subfigure}{0.99\linewidth}
        \centering
        \includegraphics[width=0.83\linewidth]{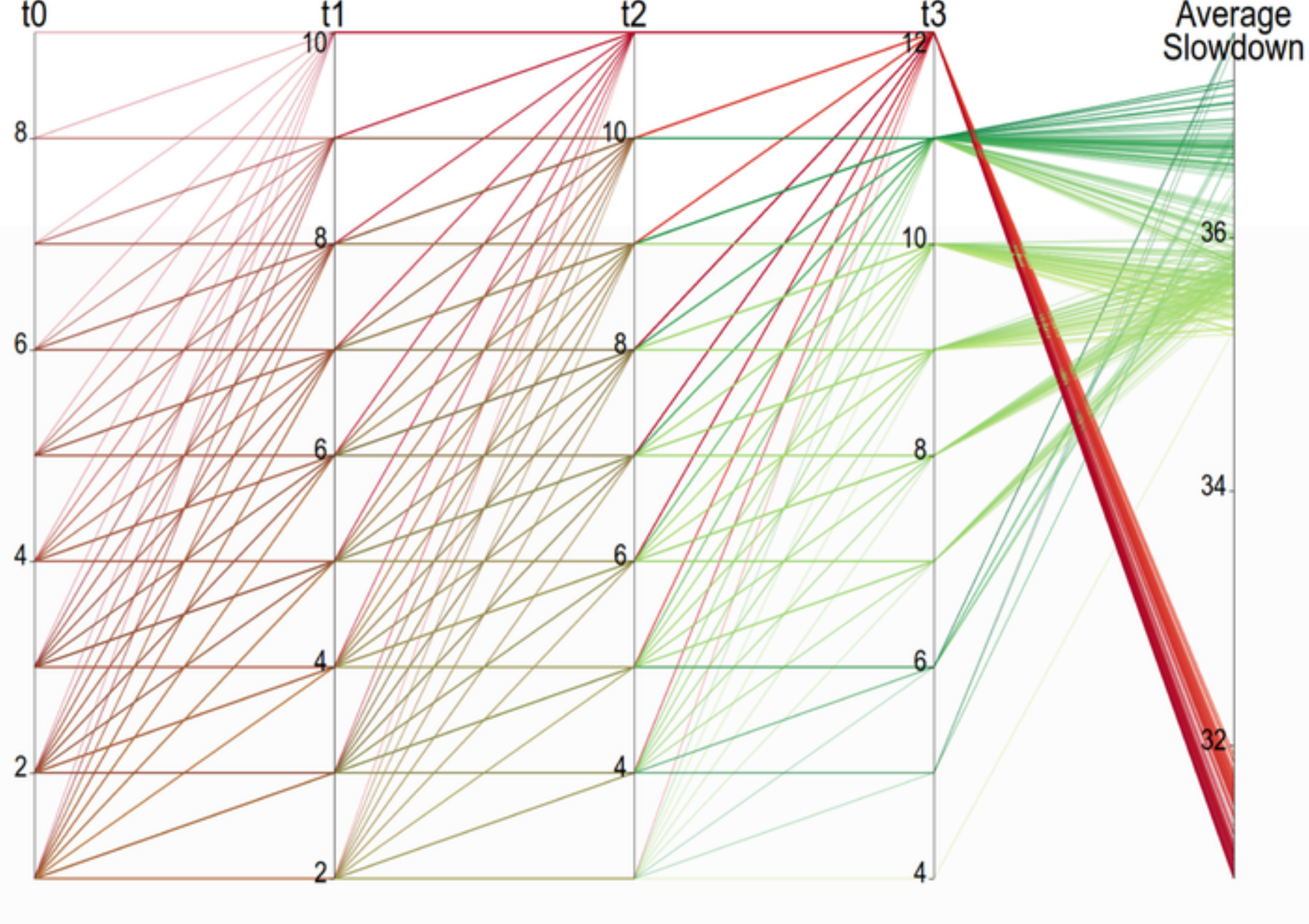}
        \caption{
            SchedSim.
        }\label{fig:schedsim_ss}
\end{subfigure}
\caption{%
      Search space examples.
  }\label{fig:searchspacede_examples}
\end{figure}


Figure~\ref{fig:searchspacede_examples} shows parallel coordinate graphs representing subsets of the search spaces of the applications. We used 20 evaluation subjects (i.e.\  an oil \& gas field, a farm, and a cluster workload) for each application. We assessed the pruning strategy for each of the 20 subjects using the other 19 as part of the knowledge base. For instance, in the evaluation of crop number 15, we used subjects $[1,14] \cup [16, 20]$ as previous experiments. For each application, we evaluated the absolute difference between the global optimum and the best-found solution using JobPruner or not. Moreover, the search space reduction is assessed for each pruned subject. We employed Particle Swarm Optimization (PSO) and Simulated Annealing (SA)~\cite{conn2009introduction} as optimization methods. Due to their stochastic nature, PSO and SA lead to different results depending on their initial conditions. We repeat the same experiment 200 times to find a 95$\%$ confidence interval of the studied metrics.

\begin{figure*}[!th]
  \begin{subfigure}{0.95\linewidth}
     \centering
     \includegraphics[width=0.95\linewidth]{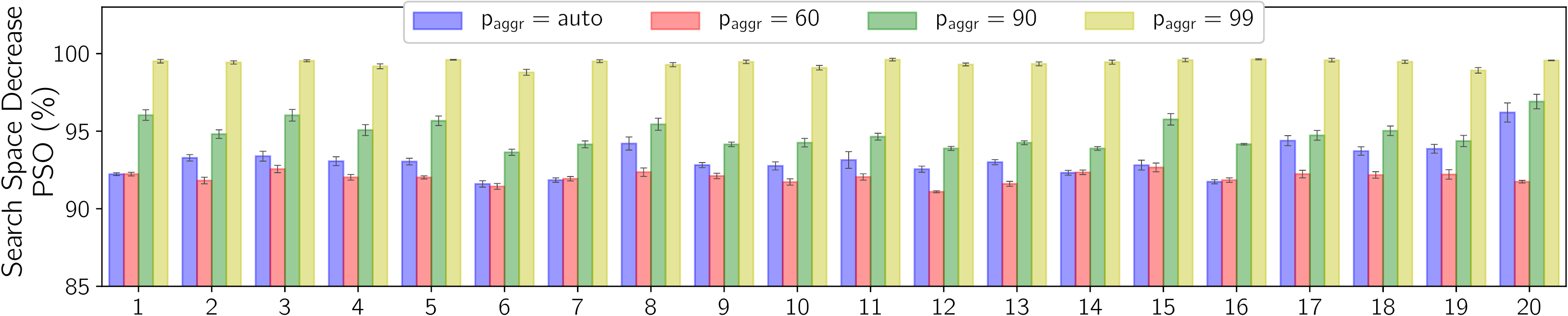}
     \includegraphics[width=0.95\linewidth]{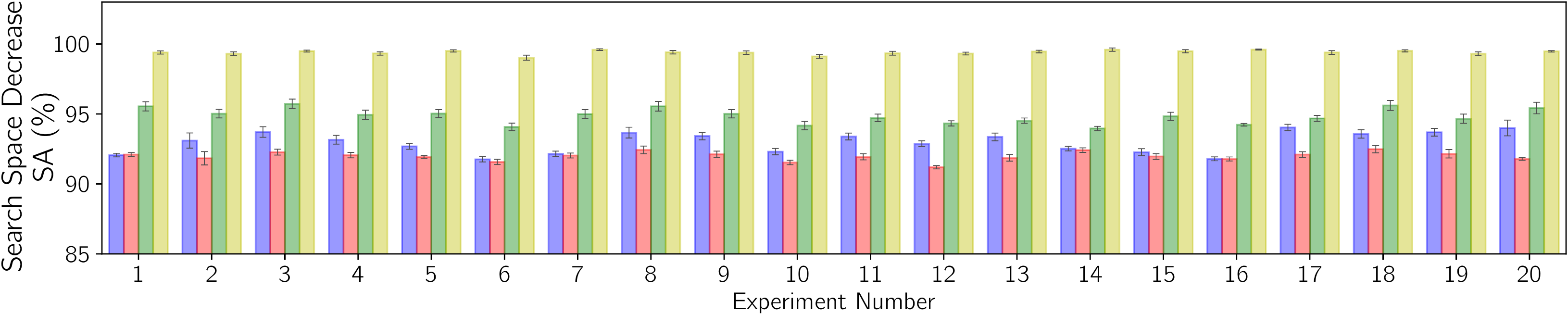}
     \caption{Seismic}\label{sfig:searchspacedecreaseseismic}
  \end{subfigure}\hfill
  \vspace{5mm}
  \begin{subfigure}{0.95\linewidth}
     \centering
     \includegraphics[width=0.95\linewidth]{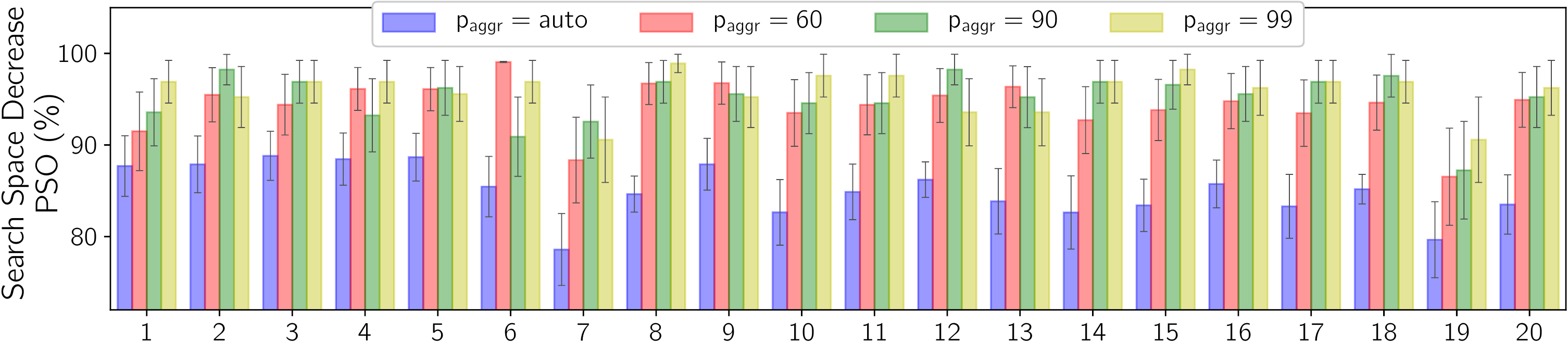}
     \includegraphics[width=0.95\linewidth]{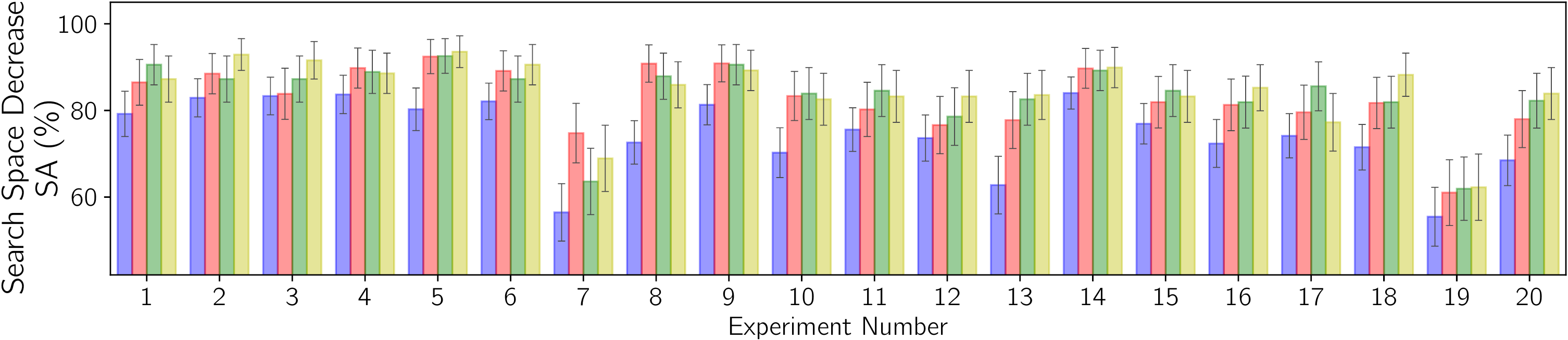}
     \caption{AgroAnalytics}\label{sfig:searchspacedecreasecrop}
  \end{subfigure}\hfill
  \vspace{5mm}
  \begin{subfigure}{0.95\linewidth}
     \centering
     \includegraphics[width=0.95\linewidth]{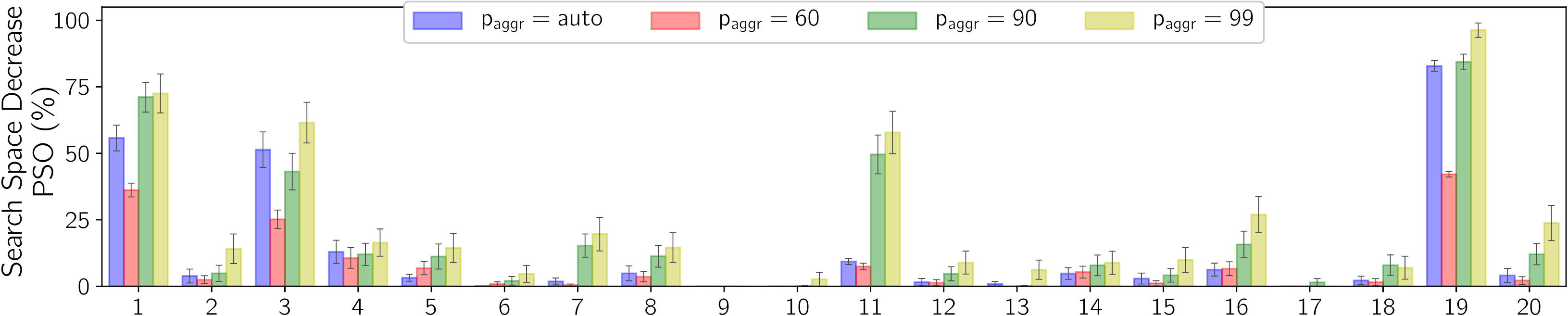}
     \includegraphics[width=0.95\linewidth]{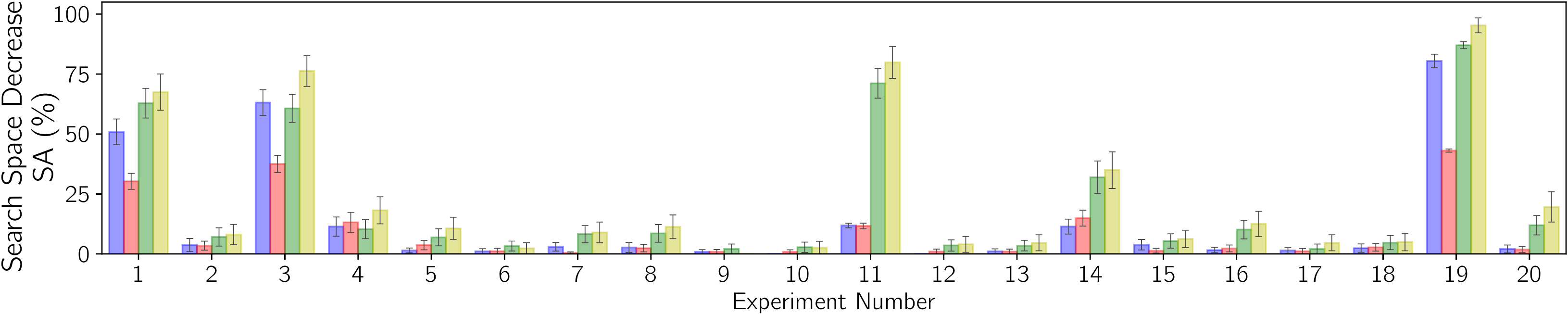}
     \caption{SchedSim}\label{sfig:sfig:searchspacedecreaseschedsim}
  \end{subfigure}\hfill
  \caption{%
      Search space decrease based on pruning aggressiveness.
  }\label{fig:searchspacedecrease}
\end{figure*}

\medskip

\noindent {\bf Heterogeneity of the experiments.} For our analysis, it is relevant to understand how similar the experiments are among themselves. Within a use case, for each experiment we calculated its normalized cross-correlation (Section~\ref{sec:selection}) against each of the other 19 experiments considering the evaluation of the entire search space. For each comparison with all other experiments, we selected the best correlation and saved it. The following values are the average of the best correlations: 0.8198, 0.8907, and 0.5756 for Seismic, AgroAnalytics, and SchedSim respectively. Therefore, we have three real world case studies with different levels of similarity.

\subsection{Results: Pruning Aggressiveness Analysis}
\label{sec:prunaggre}

Before analyzing the impact of the pruning on the quality of the results, it is important to understand how much pruning was done depending on $p_{aggr}$. For each set of experiments, we used the following values for $p_{aggr}$: 60\%, 90\%, 99\%, and auto, which means the $p_{aggr}$ was provided automatically by JobPruner (Section~\ref{sec:auto_pagg}). Figure~\ref{fig:searchspacedecrease} illustrates the search space reduction based on $p_{aggr}$ for all use cases. The behaviour of the three use cases is different, which enables us to have a more comprehensive understanding of the pruning using past experiments.  The impact of $p_{aggr}$ on pruning among the experiments depends on the shape of the search space and the correlation between experiments. Seismic and AgroAnalytics have a more steady impact of pruning on theirs experiments, whereas SchedSim is more heterogeneous and presents less cuts on search space. The search space decrease is similar when comparing PSO and SA methods, which suggests that the optimization method has little impact on pruning.

\begin{figure}[!t]
        \centering
        \includegraphics[width=1.0\linewidth]{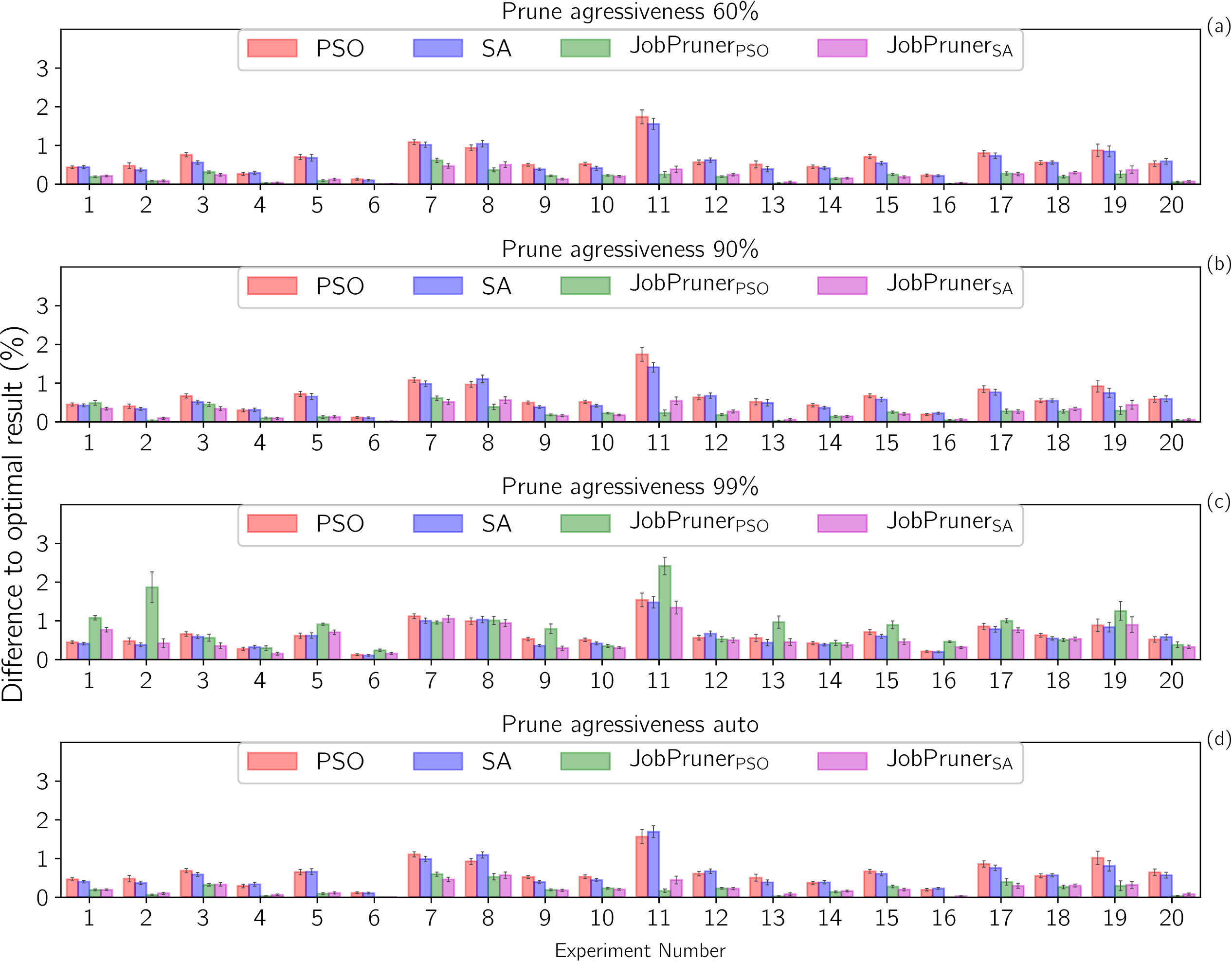}
        \caption{%
            Result quality as a function of pruning aggressiveness for Seismic.
        }\label{fig:bestimgexp}
\end{figure}
\begin{figure}[!t]
        \centering
        \includegraphics[width=1.0\linewidth]{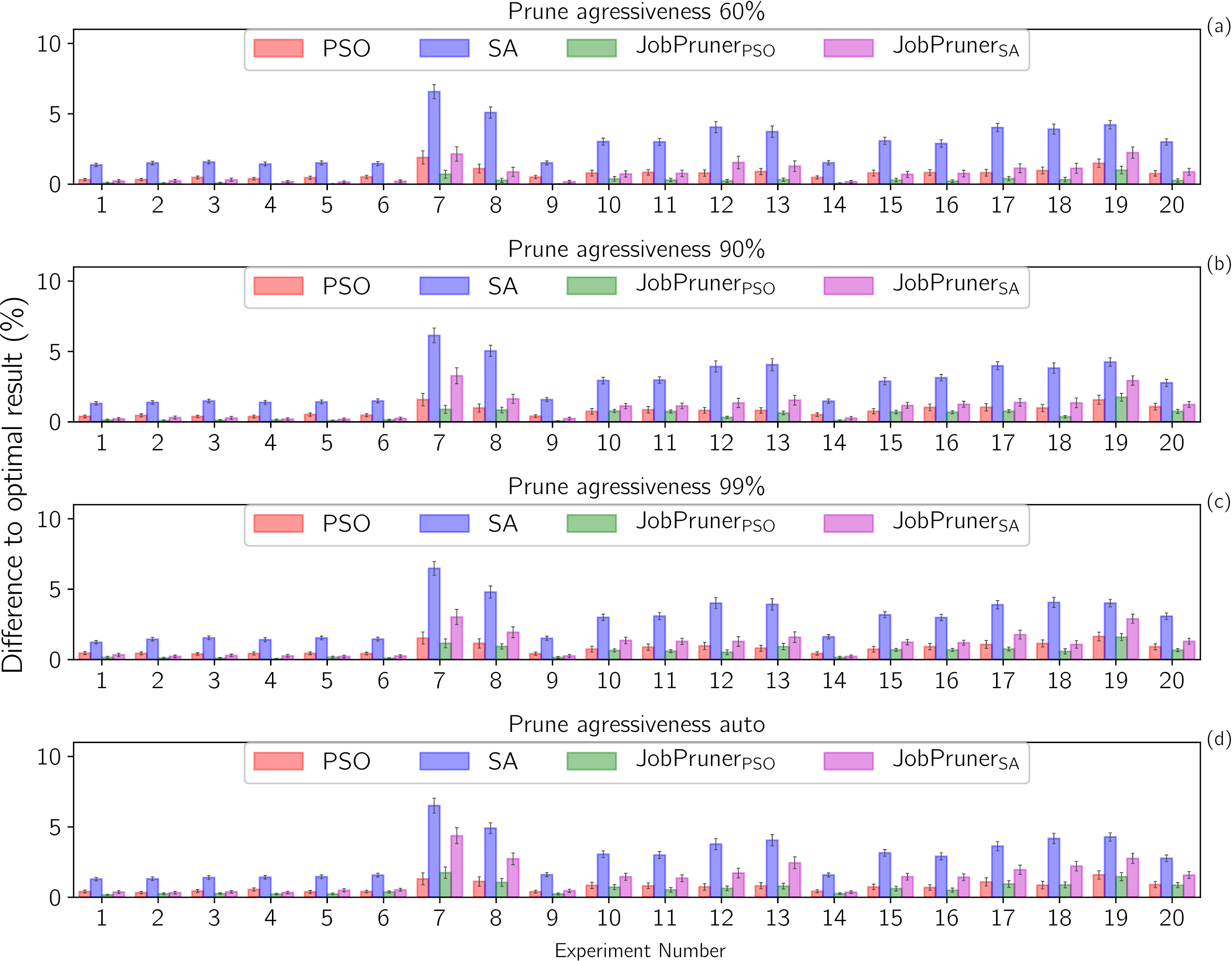}
        \caption{%
            Result quality as a function of pruning aggressiveness for AgroAnalytics.
        }\label{fig:bestcrop}
\end{figure}
\begin{figure}[!t]
        \centering
        \includegraphics[width=1.0\linewidth]{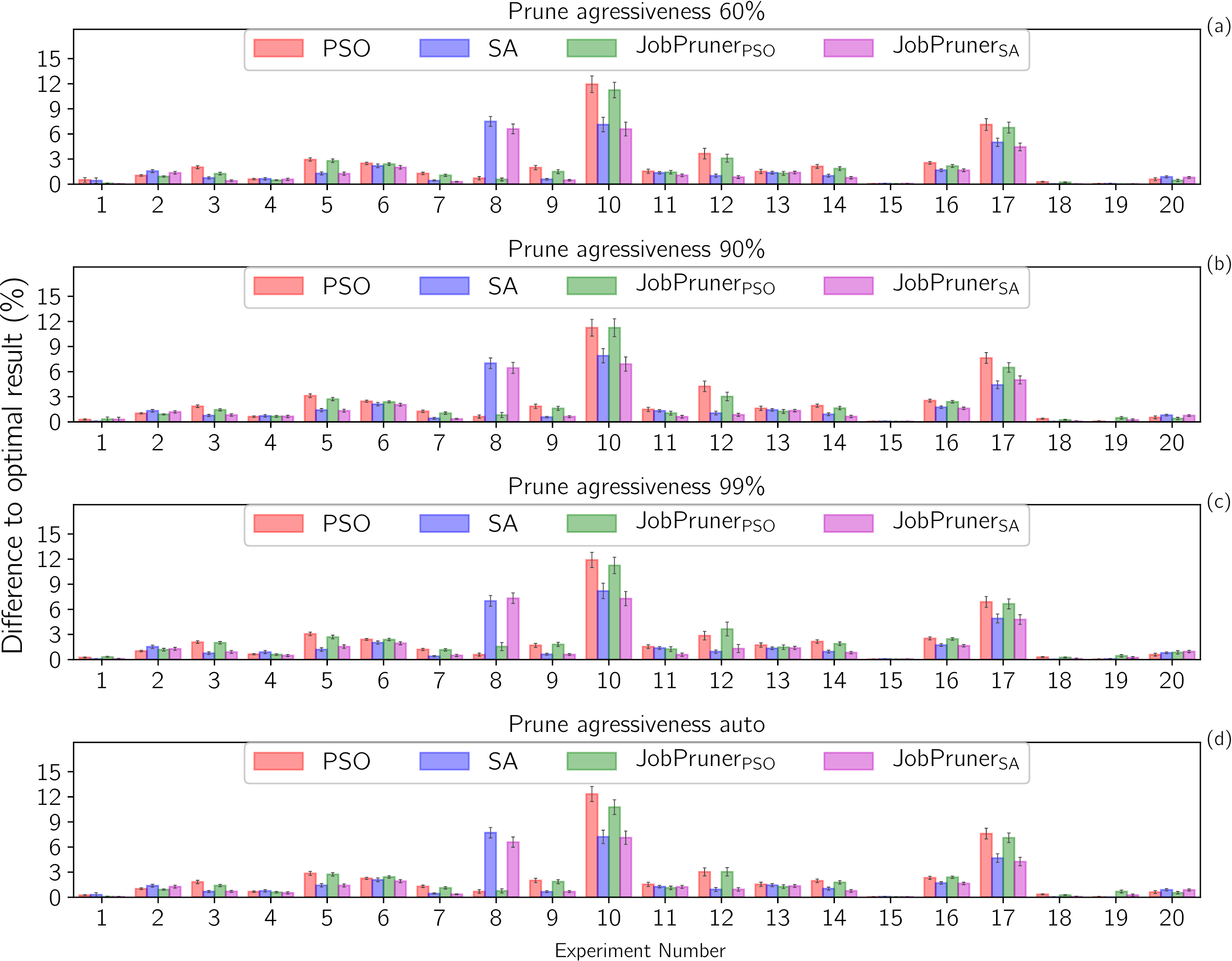}
        \caption{%
            Result quality as a function of pruning aggressiveness for SchedSim.
        }\label{fig:bestsched}
\end{figure}

Figures~\ref{fig:bestimgexp}--\ref{fig:bestsched} present a comparison of the percentage difference between the optimal result for each application subject (experiment number) using the standard optimization methods PSO and SA and the same optimization methods with search space prunes ($\mathrm{JobPruner_{PSO}}$ and $\mathrm{JobPruner_{SA}}$). The big picture of the graphs is that the higher the pruning aggressiveness the closer to optimal the results are. In general, in the worst case, pruning gets similar results to PSO/SA without pruning. Higher positive impact of pruning comes when more jobs that are far from the optimal region of the search space are eliminated, thus saving resources to more promising jobs. The cases when PSO/SA and JobPruner are similar happen when search spaces have a spiky shape and it is not possible to create surrogates that can clearly identify jobs to be pruned. Rare cases of pruning producing worse results than pure PSO also happen. This is the scenario when regions containing optimal values are not identified by jobs that were executed to identify surrogates from the knowledge base. In this case, such regions may be eliminated if pruning is too aggressive.

To exemplify, for Seismic using PSO, experiments with $p_{aggr}=60\%$, $p_{aggr}=90\%$ the difference between the best-found result and the global optimum decreases. However, for some experiments with $p_{aggr}=99\%$ (e.g., 2, 11, and 13) the difference increases when compared to $p_{aggr}=90\%$. This behaviour can be explained by analyzing the search space reduction in each experiment as illustrated in Figures~\ref{fig:bestimgexp_dbsize}--\ref{fig:bestsched_dbsize}. The search space reduces as long as the $p_{aggr}$ values increase. For $p_{aggr}=99\%$ the search space reduction gets close to 100\% which may remove the global optimal from the search space. Note that the pruning process is based on previous results and not necessarily the optimal result of one model will match the optimal of the current model. Table~\ref{tab:results_pruning} summarizes the results, observe that experiments with automatic generation of $p_{aggr}$ provide results similar to the best selected value of $p_{aggr}$.

\begin{table}[!ht]
\centering
\caption{Result quality (intervals) based on pruning aggressiveness.}
\label{tab:results_pruning}
\begin{scriptsize}
\begin{tabular}{@{}llcccc@{}}
\toprule
    Opt & $p_{aggr}$ &   \begin{tabular}[c]{@{}l@{}}\shortstack{\% diff \\ (Seismic)}\end{tabular} & \begin{tabular}[c]{@{}l@{}}\shortstack{\% diff \\ (AgroAnalytics)}\end{tabular} & \begin{tabular}[c]{@{}l@{}}\shortstack{\% diff \\ (SchedSim)}\end{tabular} \\ \midrule
	& 0 	& [0.571, 0.705] & [0.580, 0.970] & [2.181, 2.212] \\
	& 60 	& [0.158, 0.219] & [0.145, 0.359] & [1.943, 1.979] \\
PSO & 90 	& [0.179, 0.257] & [0.395, 0.609] & [1.941, 2.030] \\
	& 99	& [0.745, 0.947] & [0.415, 0.658] & [2.120, 2.224] \\
	& auto 	& [0.168, 0.239] & [0.481, 0.793] & [2.006, 2.044] \\
	\midrule
	& 0		& [0.530, 0.652] & [2.653, 3.148] & [1.675, 1.752] \\
	& 60 	& [0.168, 0.238] & [0.558, 1.024] & [1.478, 1.552] \\
SA 	& 90 	& [0.198, 0.282] & [0.848, 1.279] & [1.517, 1.627] \\
	& 99	& [0.484, 0.630] & [0.884, 1.320] & [1.605, 1.736] \\
	& auto 	& [0.178, 0.256] & [1.198, 1.677] & [1.538, 1.628] \\
\bottomrule
\end{tabular}
\end{scriptsize}
\end{table}

\begin{figure*}[!h]
  \begin{subfigure}{0.95\textwidth}
     \centering
     \includegraphics[width=0.95\linewidth]{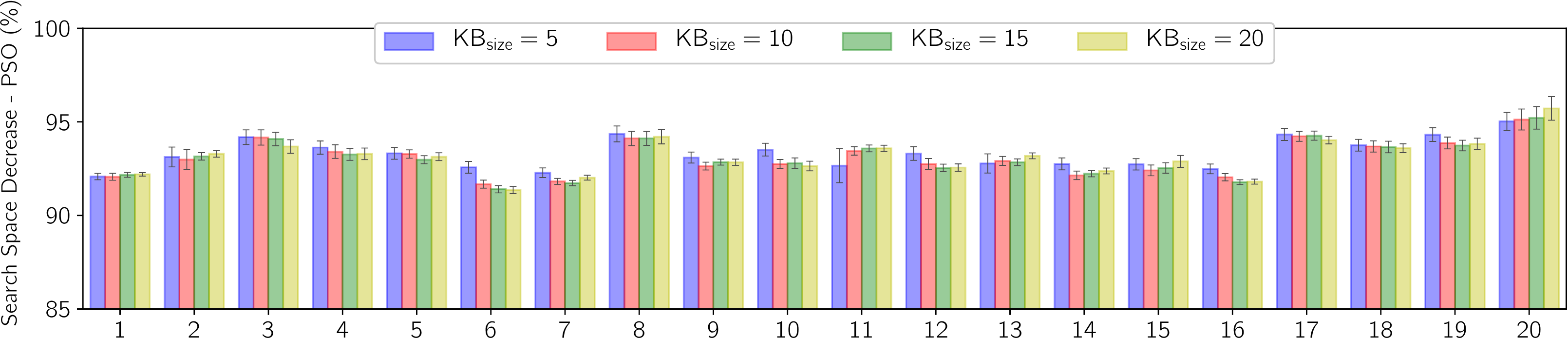}
     \includegraphics[width=0.95\linewidth]{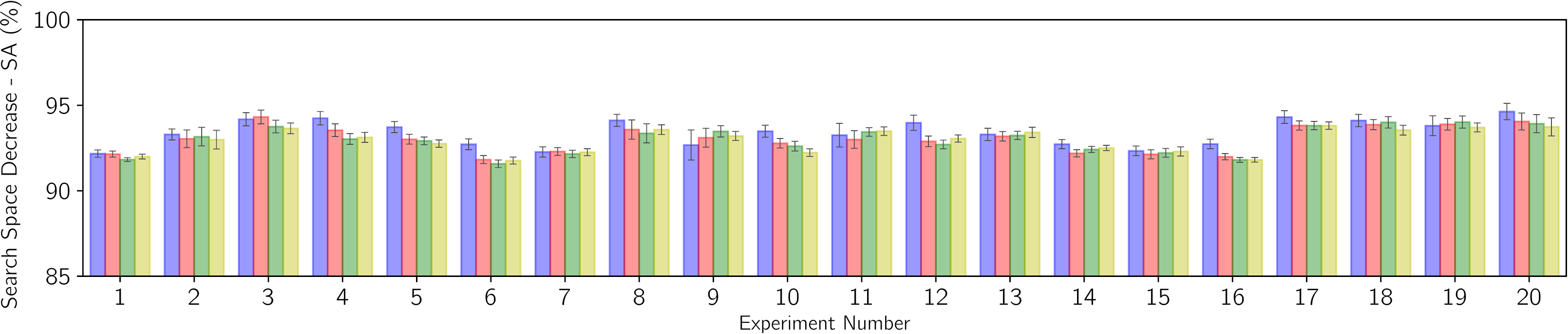}
     \caption{Seismic}\label{sfig:searchspacedecreaseseismic_db}
  \end{subfigure}\hfill
  \vspace{5mm}
  \begin{subfigure}{0.95\textwidth}
     \centering
     \includegraphics[width=0.95\linewidth]{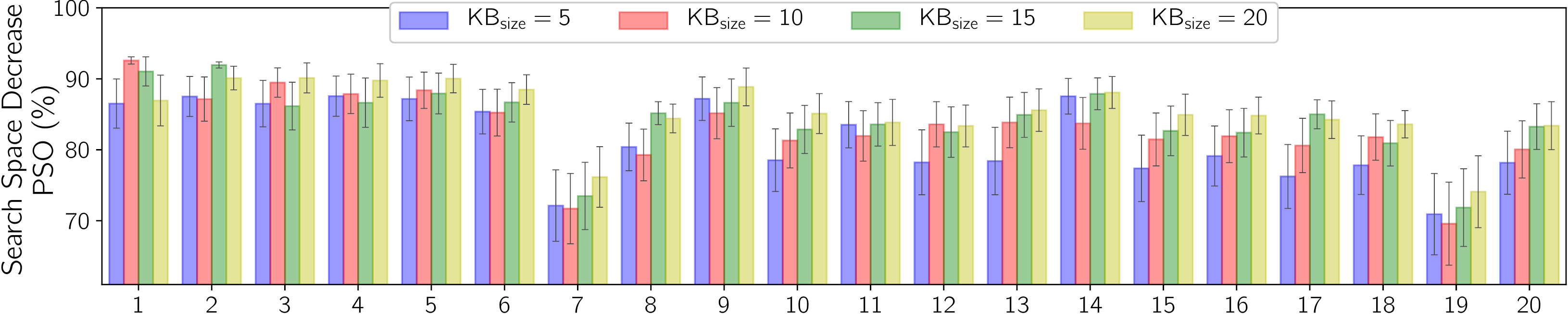}
     \includegraphics[width=0.95\linewidth]{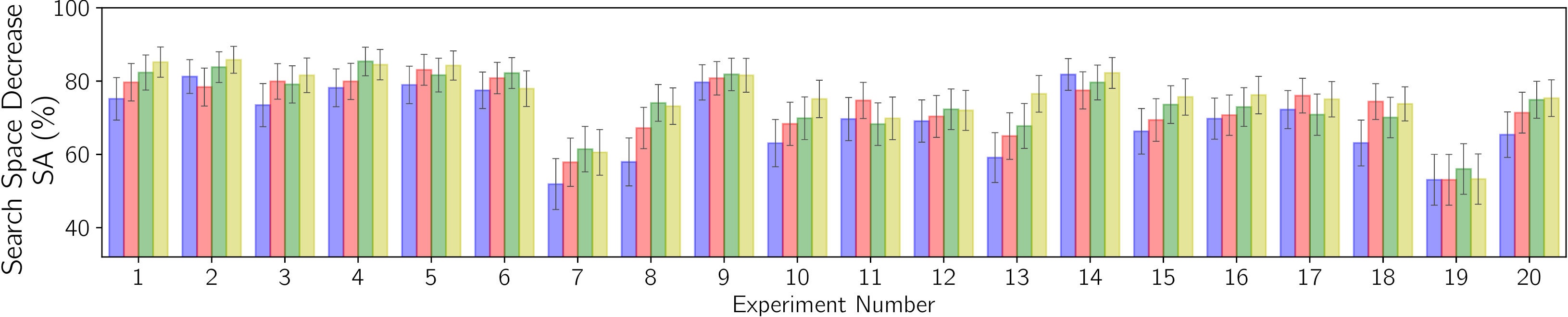}
     \caption{AgroAnalytics}\label{sfig:searchspacedecreasecrop_db}
  \end{subfigure}\hfill
  \vspace{5mm}
  \begin{subfigure}{0.95\textwidth}
     \centering
     \includegraphics[width=0.95\linewidth]{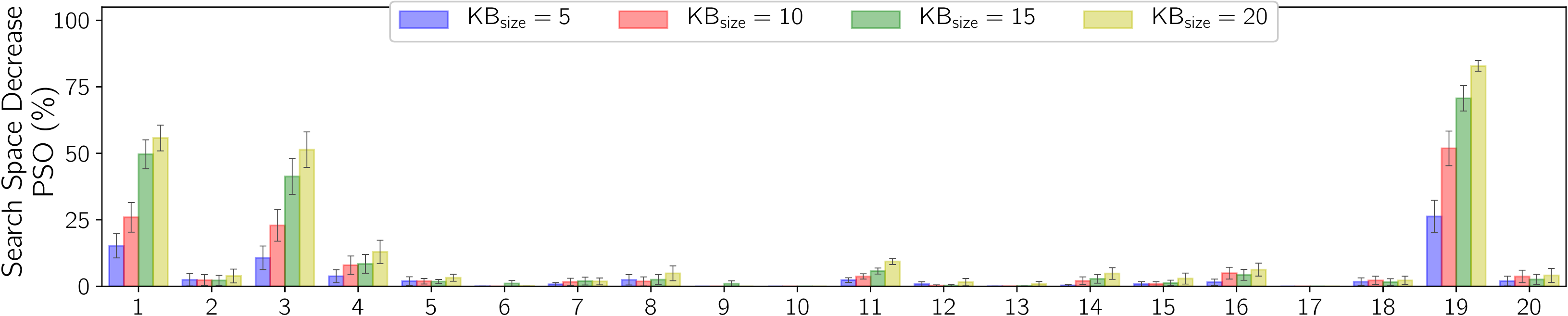}
     \includegraphics[width=0.95\linewidth]{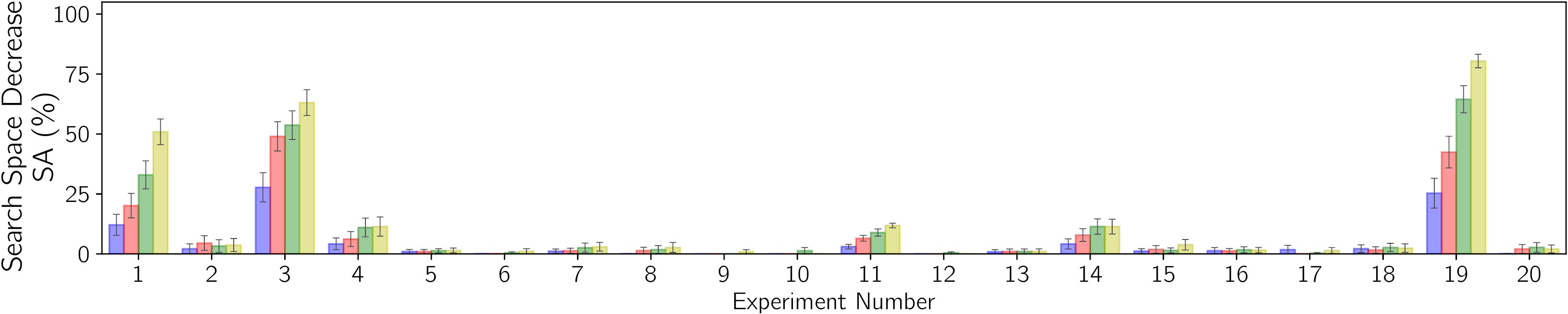}
     \caption{SchedSim}\label{sfig:searchspacedecreaseschedsim_db}
  \end{subfigure}\hfill
  \caption{%
      Search space decrease for different previous experiments knowledge base sizes.
  }\label{fig:searchspacedecrease_dbsize}
\end{figure*}

\subsection{Results: Previous Experiments Database Size Analysis}
\label{sec:prevdbsize}

An important aspect to be evaluated is the impact of knowledge base size on the search space reduction.  It is intuitive that the larger the knowledge base the more possibilities for pruning JobPruner has. However, as illustrated in Figure~\ref{fig:searchspacedecrease_dbsize}, use cases have different behaviour depending on the knowledge base size. In the following study we changed the knowledge base size of previous experiments and used automatic generation of prune aggressiveness $(p_{aggr}=auto)$. For Seismic, we observe that influence is minimal due to the high similarities of the search space shapes among different experiments. AgroAnalytics and SchedSim show a more heterogeneous behaviour. The shapes of the search spaces vary considerably among different experiments, especially for SchedSim. The reduction in search spaces varies according to experiment number, but in general the cuts in search space increase with the knowledge base size.


\begin{figure}[!t]
        \centering
        \includegraphics[width=1.0\linewidth]{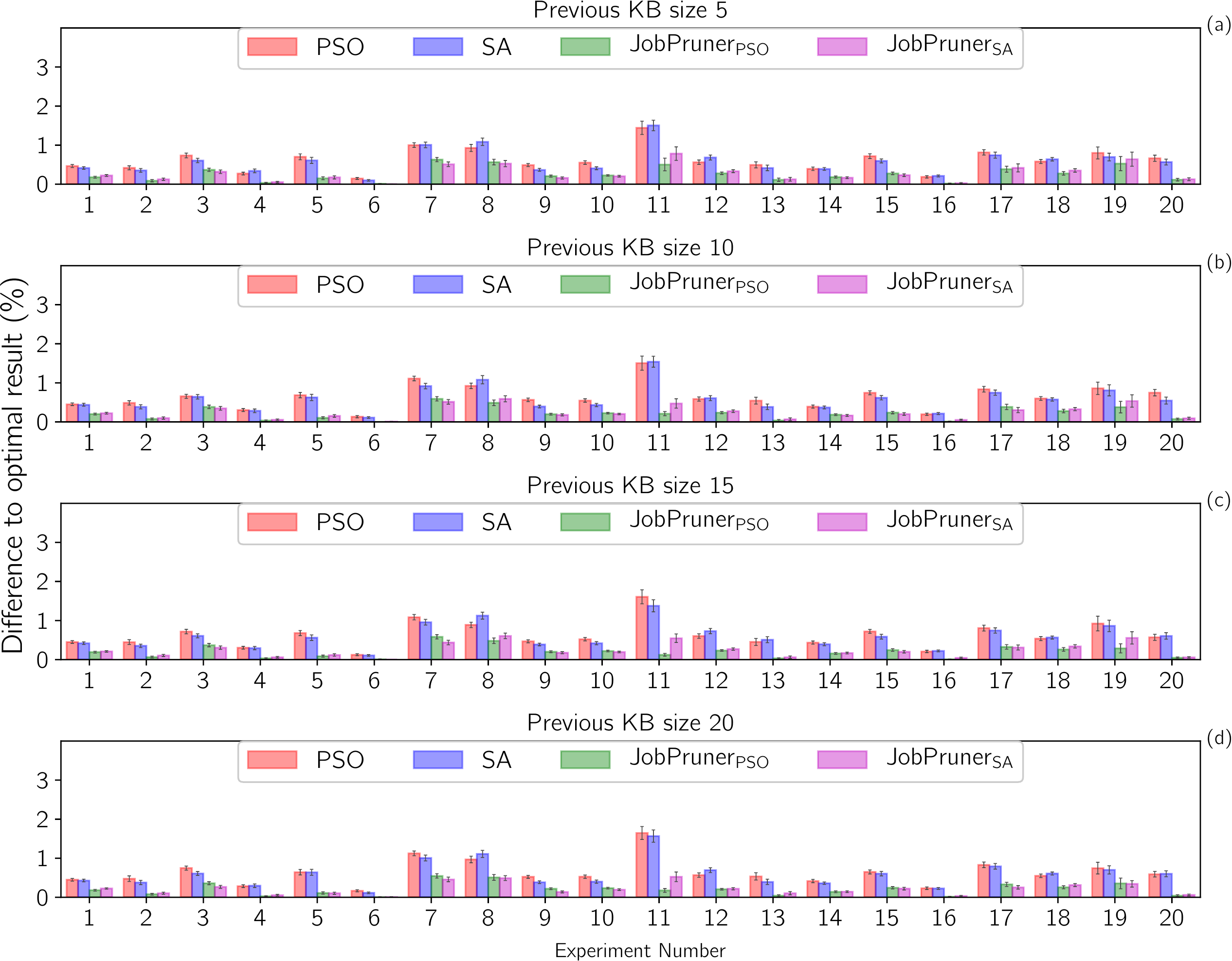}
        \caption{%
            Result quality as a function of the knowledge base size for Seismic.
        }\label{fig:bestimgexp_dbsize}
\end{figure}
\begin{figure}[!t]
        \centering
        \includegraphics[width=1.0\linewidth]{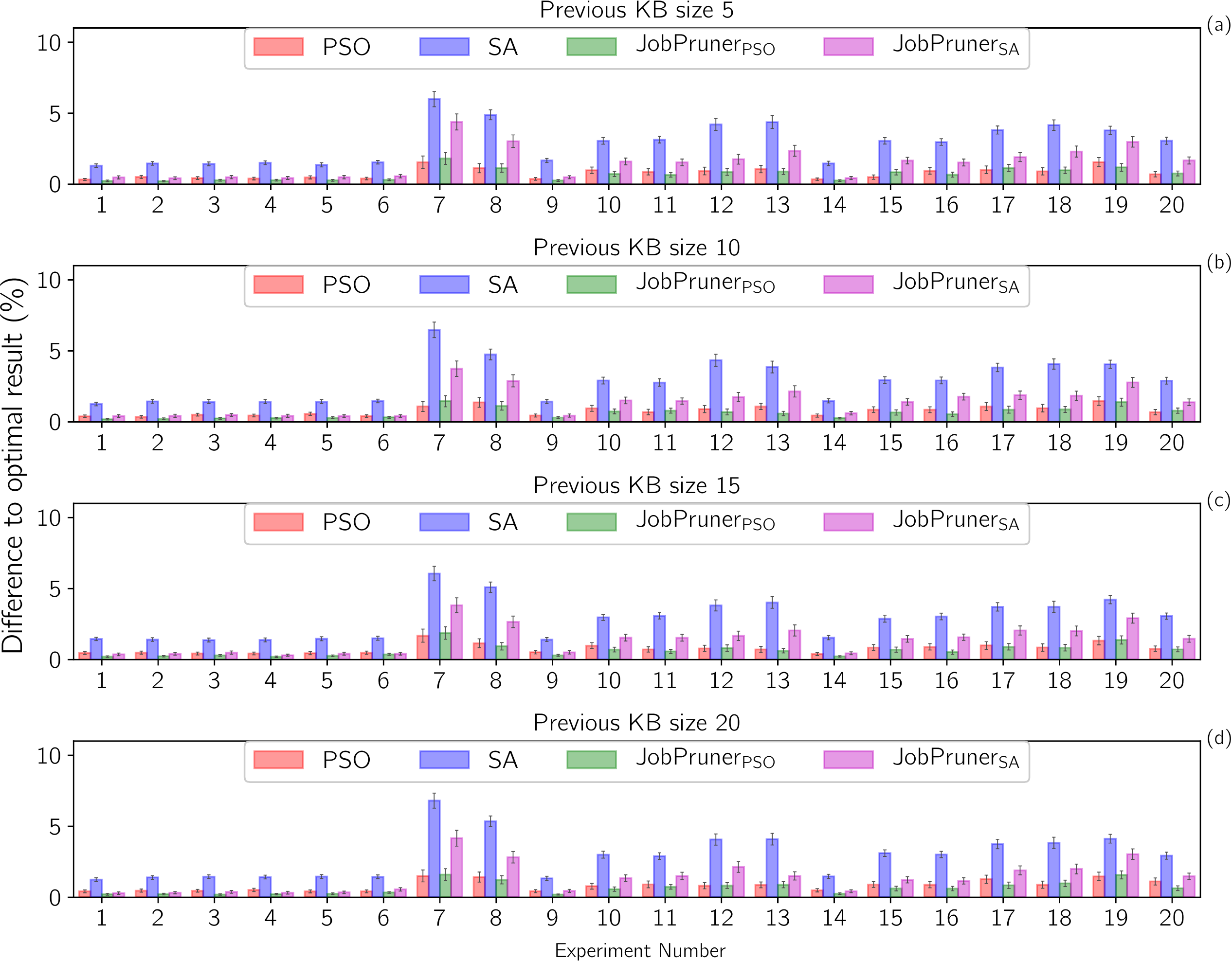}
        \caption{%
            Result quality as a function of the knowledge base size for AgroAnalytics.
        }\label{fig:bestcrop_dbsize}
\end{figure}
\begin{figure}[!t]
        \centering
        \includegraphics[width=1.0\linewidth]{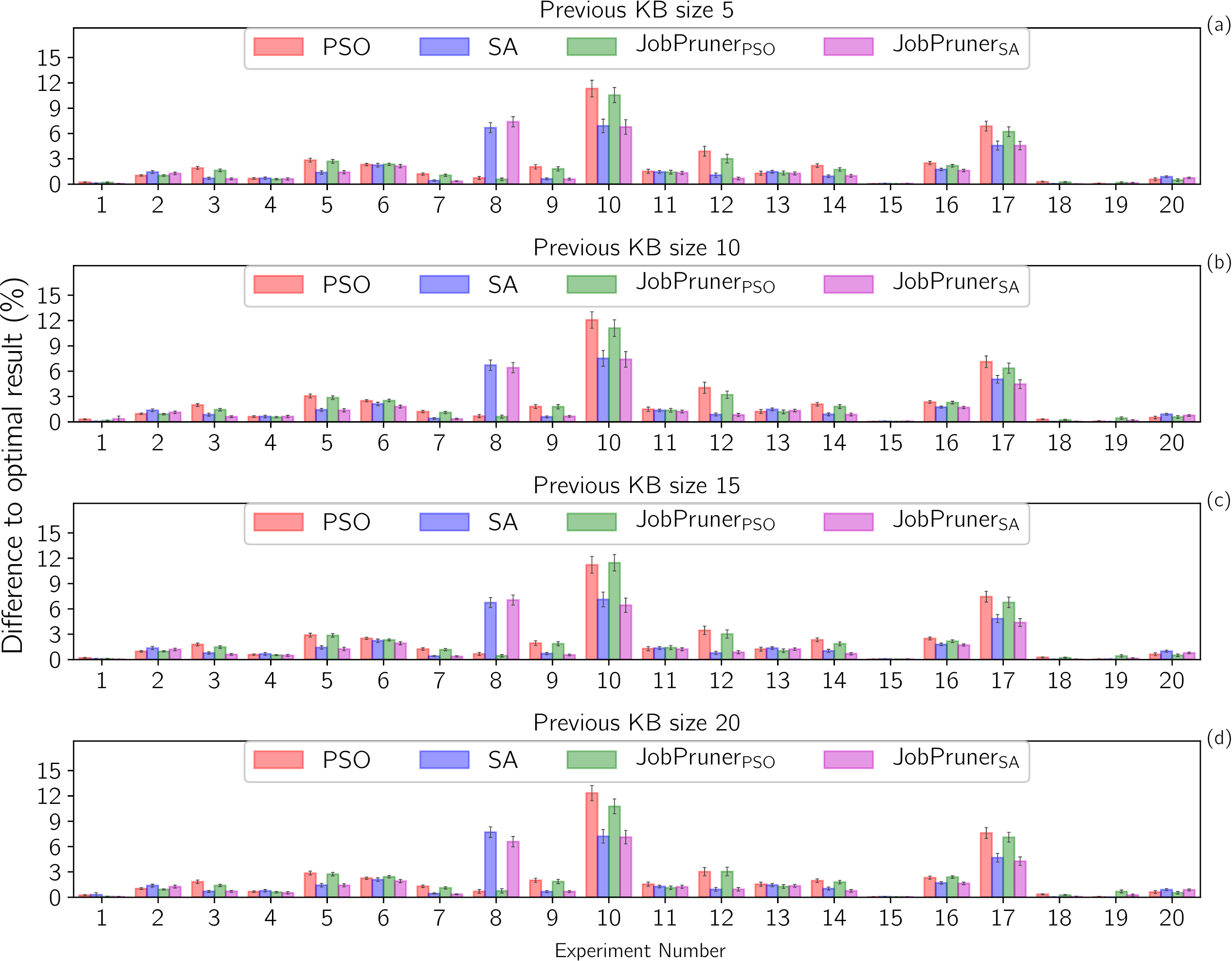}
        \caption{%
            Result quality as a function of the knowledge base size for SchedSim.\vspace{5mm}
        }\label{fig:bestsched_dbsize}
\end{figure}

Figures~\ref{fig:bestimgexp_dbsize},~\ref{fig:bestcrop_dbsize},~\ref{fig:bestsched_dbsize} present the quality of the results as a function of the knowledge base size for the three use cases. For Seismic we observe considerable improvements when compare knowledge base (KB) 5 to the others (10, 15, and 20). However, the experiments with size 10 to 20 present very similar results. This behaviour can be explained by the small variation in search size pruning due to high similarity between Seismic experiments.

For AgroAnalytics, we also observe low variation of the results among different KB sizes. This happens due to the low variation of search space cuts (Figure~\ref{sfig:searchspacedecreasecrop_db}). For SchedSim, improvements due to DB size are not clearly observable. For this use case, even though there is a reduction in the search space due to a more comprehensive knowledge base (Figure~\ref{sfig:searchspacedecreaseschedsim_db}), this reduction is not significant enough to generate spare computing power to reach for better optimization results due to the great variability in the search space among the experiments of SchedSim. Table~\ref{tab:results_dbsize} summarizes the results. For all experiments, there is an improvement when comparing experiments with and without KB, and, for some, this improvement continues when the KB increases along all experiments (i.e., Seismic and AgroAnalytics with SA).

\begin{table}[!ht]
\centering
\caption{Result quality (intervals) based on knowledge base size.}
\label{tab:results_dbsize}
\begin{scriptsize}
\begin{tabular}{@{}llccc@{}}
\toprule
Optimizer & \begin{tabular}[c]{@{}l@{}}Previous\\ Exp size\end{tabular} & \begin{tabular}[c]{@{}l@{}}\shortstack{\% diff \\ (Seismic)}\end{tabular} & \begin{tabular}[c]{@{}l@{}}\shortstack{\% diff \\ (AgroAnalytics)}\end{tabular} & \begin{tabular}[c]{@{}l@{}}\shortstack{\% diff \\ (SchedSim)}\end{tabular} \\ \midrule
	& 0 & [0.565,  0.696] &  [0.587,  0.980]	& [2.157,  2.188]	\\
	& 5 & [0.208,  0.305] &  [0.523,  0.850] & [1.920,  1.972] \\
PSO	& 10 & [0.178,  0.254] & [0.472,  0.786] &[1.982,  2.034] \\
	& 15 & [0.163,  0.231] & [0.472,  0.786] & [1.995,  2.033]\\
	& 20 & [0.168,  0.238] & [0.494,  0.804] & [2.006,  2.044]\\
	\midrule
	& 0 & [0.527,  0.652] & [2.636,  3.133] &[1.644,  1.720] \\
	& 5 & [0.222,  0.326] & [1.280,  1.769] &[1.575,  1.651] \\
SA	& 10 & [0.197,  0.287] & [1.175,  1.642] &[1.531,  1.646]\\
	& 15 & [0.195,  0.279] & [1.171,  1.636] &[1.496,  1.573] \\
	& 20 & [0.174,  0.250] & [1.135,  1.597]&[1.538,  1.628] \\
\bottomrule
\end{tabular}
\end{scriptsize}
\end{table}

\section{Conclusion}

Scientists and engineers always look for ways to make their models more realistic to have insights about their subject of study. This usually imposes evaluation of a large number of computational jobs executed on HPC machines. In this paper we investigated the possibility and benefits of using data from past experiments to help identify unnecessary jobs to be executed and speed up experiments of these professionals. We also introduced a machine learning-based tool, called JobPruner, to automate the process of identifying such jobs.

We executed a series of experiments using three real use case applications from different fields and were able to draw the following lessons:
\begin{itemize}
	\item Pruning aggressiveness and search space sizes: Search space shapes have high influence on selecting jobs to be eliminated. Spiky search spaces tend to be more difficult to create surrogates that can easily identify jobs to be pruned;
	\item Knowledge base size: when experiments contain similar shapes of search spaces, expanding knowledge base size has no impact on pruning and quality of results. However, for experiments that are more heterogeneous, any additional experiments added to the knowledge base can bring new insights to prune jobs;
	\item Experiment similarities: even when experiments have low correlation, it is still possible learn portions of the search space that will not bring value to the user and the aggressiveness to make the proper prunes can be automatically identified based on surrogates generated from previous experiments.
\end{itemize}

This work is an indication that the reuse of experiments of the same kind of object/model is possible and we envision this is a rich research direction of exploring big data for optimizing HPC scientific experiments.

\balance
\bibliographystyle{elsarticle-num}
\bibliography{references}

\end{document}